\documentclass[journal]{IEEEtran}%TWC

\usepackage{amsmath}
\usepackage{bm}
\usepackage{amsthm}
\usepackage{graphicx}
\usepackage{float}
\usepackage{subfig}
\usepackage{cite}
\usepackage{enumerate}
\usepackage{mathrsfs}
\usepackage{multirow}
\usepackage{amsfonts}
\usepackage{makecell}
\usepackage{url}
\usepackage{CJK}
\usepackage{indentfirst}
\usepackage{algorithm}
\usepackage{algpseudocode}
\usepackage{color, soul}
\usepackage{indentfirst} 

\begin{document}
\title{\huge{Towards Ubiquitous Sensing and Localization With Reconfigurable Intelligent Surfaces}}

\author{
	\IEEEauthorblockN{
		{Hongliang Zhang}, \IEEEmembership{Member, IEEE},
		{Boya Di}, \IEEEmembership{Member, IEEE},
		{Kaigui Bian}, \IEEEmembership{Senior Member, IEEE},\\
		{Zhu Han}, \IEEEmembership{Fellow, IEEE},
		{H. Vincent Poor}, \IEEEmembership{Life Fellow, IEEE}},
		{and Lingyang Song}, \IEEEmembership{Fellow, IEEE}\\

	\thanks{H. Zhang is with Department of Electronics, Peking University, Beijing, China, and also with Department of Electrical and Computer Engineering, Princeton University, Princeton, NJ, USA.} 
		
	\thanks{B. Di and L. Song are with Department of Electronics, Peking University, Beijing, China.}
	
	\thanks{K. Bian is with Department of Computer Science, Peking University, Beijing, China.}
	
	\thanks{Z. Han is with with Electrical and Computer Engineering Department, University of Houston, Houston, TX, USA, and also with the Department of Computer Science and Engineering, Kyung Hee University, Seoul, South Korea.}
	
	\thanks{H. V. Poor is with Department of Electrical and Computer Engineering, Princeton University, Princeton, NJ, USA.}
}

\maketitle

\begin{abstract}
In future cellular systems, wireless localization and sensing functions will be built-in for specific applications, e.g., navigation, transportation, and healthcare, and to support flexible and seamless connectivity. Driven by this trend, the need rises for fine-resolution sensing solutions and cm-level localization accuracy, while the accuracy of current wireless systems is limited by the quality of the propagation environment. Recently, with the development of new materials, reconfigurable intelligent surfaces (RISs) provide an opportunity to reshape and control the electromagnetic characteristics of the environment, which can be utilized to improve the performance of wireless sensing and localization. In this tutorial, we will first review the background and motivation to utilize wireless signals for sensing and localization. Next, we introduce how to incorporate RIS into applications of sensing and localization, including key challenges and enabling techniques, and then some case studies will be presented. Finally, future research directions will also be discussed.
\end{abstract}


\section{Introduction}%\vspace{-1mm}
                                   
\subsection{Sensing and Localization in Future Cellular Systems: Basic Requirements}
Driven by a wide range of emerging applications such as automated vehicles and robots, it requires future wireless networks to enable high-resolution environmental awareness in order to fulfill interactions between digital and physical worlds. This can be achieved by sensing and localization functions in wireless networks. In other words, a device in the wireless network should have the ability to know its location as well as detect the presence of objects, their shape, location, and speed of movements in the operating environment using transmitted or received radio signals \cite{TVH-2021}. 

In this regard, the fifth generation (5G) networks already provide possibilities for accurate localization and sensing services with its larger bandwidth and massive antenna array, and the sixth generation (6G) will continue this trend~\cite{AABCDDEGHHJJMMMMRTTY-2020}. As defined by the 3GPP, 5G needs to satisfy a level of accuracy less than 1 meter for more than 95\% of network area~\cite{3GPP-2018}, and it is envisioned to be achieve sub-centimeter accuracy in 6G \cite{X-2021}. These requires evolution of radio frequency (RF) sensing and localization techniques to support such high accuracy. 

\subsection{Motivations: Why RISs?}
Current wireless systems highly depend on the quality of the propagation environment, which is conventionally modeled as an exogenous entity that can only be adapted to but not be controlled. This is challenged by future wireless systems which integrate communication, sensing, and localization functions into one single platform. It is expected that the wireless environment should be treated as part of the network design parameters~\cite{HBLZ-2021}.

With the development of meta-materials, this new design of wireless networks is facilitated by an emerging technology which is referred to as Reconfigurable Intelligent Surfaces~(RISs)~\cite{MAMMCJS-2020}. The RIS is thin layers of meta-materials capable of shaping wireless signals that impinge upon the surface such that propagation environments can be customized to fulfill specific system requirements \cite{ZMWWMHS-2022}. This can be achieved by controlling the phases and amplitudes of the impinging radio signals through nearly passive and low-cost elements embedded on the RIS \cite{MHLKZG-2020}. As a result, the RIS can provide a favorable propagation condition to improve the sensing and localization accuracy. Specially, the RIS can manipulate the signals from different targets or locations to be more distinguishable, which makes the receiver easier to detect the targets or estimate the locations.

\subsection{Use Cases}
Wireless sensing and localization have a variety of applications in our daily lives, such as indoor navigation, transportation, healthcare, and security, which contribute to the development of a smart city \cite{FMKMMM-2020}. In the following, we will elaborate these application scenarios to show the importance of accurate and ubiquitous sensing and localization.
\begin{itemize}
	\item \textbf{Indoor Navigation:} Indoor navigation is an important use case in shopping malls, factories, and airports, which requires to provide an accurate localization for the user. Different from outdoor scenarios, indoor localization suffers from severe LoS blockage that will significantly degrade the accuracy \cite{KPEYCFKDAAM-2016}. This can be alleviated by deploying an RIS on the wall to provide virtual LoS links, which is particularly important in industrial Internet-of-Things applications \cite{EMOSAGMKYM-2018}, e.g., factory robots.
	
	\item \textbf{Intelligent Transportation:} Autonomous driving \cite{BMSDE-2016} and vehicle-to-everything (V2X) communications \cite{PBHKL-2018} are envisioned as two potential lines to realize intelligent transportation. For autonomous driving, it is critical to build a real-time map and detect its environment, which requires to have accurate distances among vehicles or between the vehicle and surrounding obstacles for safe operations. For V2X communications, the measurement of the velocity can help predict the location of a vehicle, which will further improve the communication performance. With the RIS attached on buildings and billboards, localization and sensing accuracy can be further enhanced, and thus improving the safety and efficiency of Transportation systems.
	
	\item \textbf{Healthcare:} Physical activity recognition for healthcare such as fall detection is an important application of wireless sensing due to its contact-free nature, i.e., the users do not need to carry devices or modify their daily routine~\cite{HDYJYS-2017}. For these applications, the reliability and accuracy are most important concerns, which are highly dependent on the quality of channel conditions~\cite{SF-2019}. By deploying some RISs, the wireless propagation environment can be customized, thus alleviating these issues.
		
	\item \textbf{Security:} RF sensing can also be used for security issues, e.g., theft monitoring \cite{JQMY-2018}. These applications require a extremely high sensing accuracy, as missing detection might cause serious consequences. However, the accuracy of traditional sensing methods is limited by channel conditions. The accuracy can be further improved by deploying the RIS to provide favorable channel conditions.
\end{itemize}

\subsection{Contribution and Organization}
In this paper, we aim to provide a tutorial overview on achieving RF sensing and localization using RISs, by reviewing the state-of-the-art results in the literature, presenting new ideas to solve the main challenges for sensing and localization accuracy improvement, and introducing hardware prototype implementation. Moreover, we identify promising research directions related to RIS-aided RF sensing and localization for motivating future work. It is worth noting that to the authors' best knowledge, this paper is the first tutorial paper to address the issues in RF sensing and localization applications with RISs. 

The rest of this paper is organized as follows. We present the fundamentals in Section \ref{Fundamentals}, including the principle of RF sensing and localization, as well as the basics of RISs and corresponding signal models. Sections \ref{sensing} and \ref{localization} elaborate on the enabling technologies for RIS-aided sensing and localization applications, respectively. In Section \ref{Implementation}, we introduce how to implement such a system and show some important experimental results. In Section \ref{Future}, we outline other possible future directions. Finally, we conclude this paper in Section \ref{Conclusion}.

\section{Fundamentals of RF Sensing and Localization with RISs}
\label{Fundamentals}
In this section, we provide some preliminary backgrounds on RF sensing and localization as well as RIS technology. In Section \ref{sub:working principle}, we first introduce how to use RF signals to realize sensing and localization. In Section \ref{sub:RIS}, we then present some basics of RIS, and finally in Section \ref{sub:RF signal}, we show the RF signal models in an RIS-aided wireless network. 

\subsection{Working Principle}
\label{sub:working principle}

The basic idea to utilize RF signals for sensing and localization applications is to extract information from wireless signals. However, there are still some differences between sensing and localization applications, which will be elaborated below.

\subsubsection{RF Sensing}
\begin{figure}[!t]
	\centering
	\includegraphics[width=3.2in]{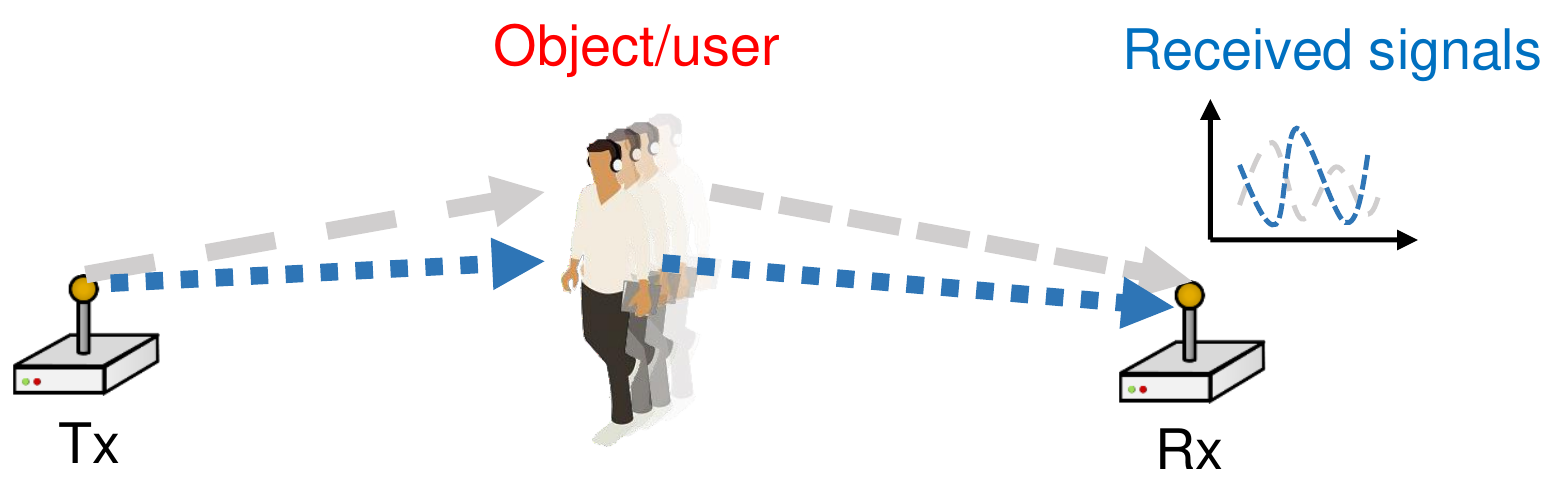}
	\caption{Illustration of sensing using wireless signals.}
	\label{sensing}
\end{figure}

The working principle underlying RF sensing is that the presence of objects will cause changes in wireless signals, which results in a variation of certain properties of received signals \cite{JHYYC-2020}. As a result, we can detect the existence of objects from the variation of received signals. A typical RF sensing system is shown in Fig. \ref{sensing}. There exists a pair of the transmitter (Tx) and the receiver (Rx), and the Rx will analyze the received signals to recognize the movement of users or the existence of objects.

\subsubsection{RF Localization}
\begin{figure}[!t]
	\centering
	\includegraphics[width=2.5in]{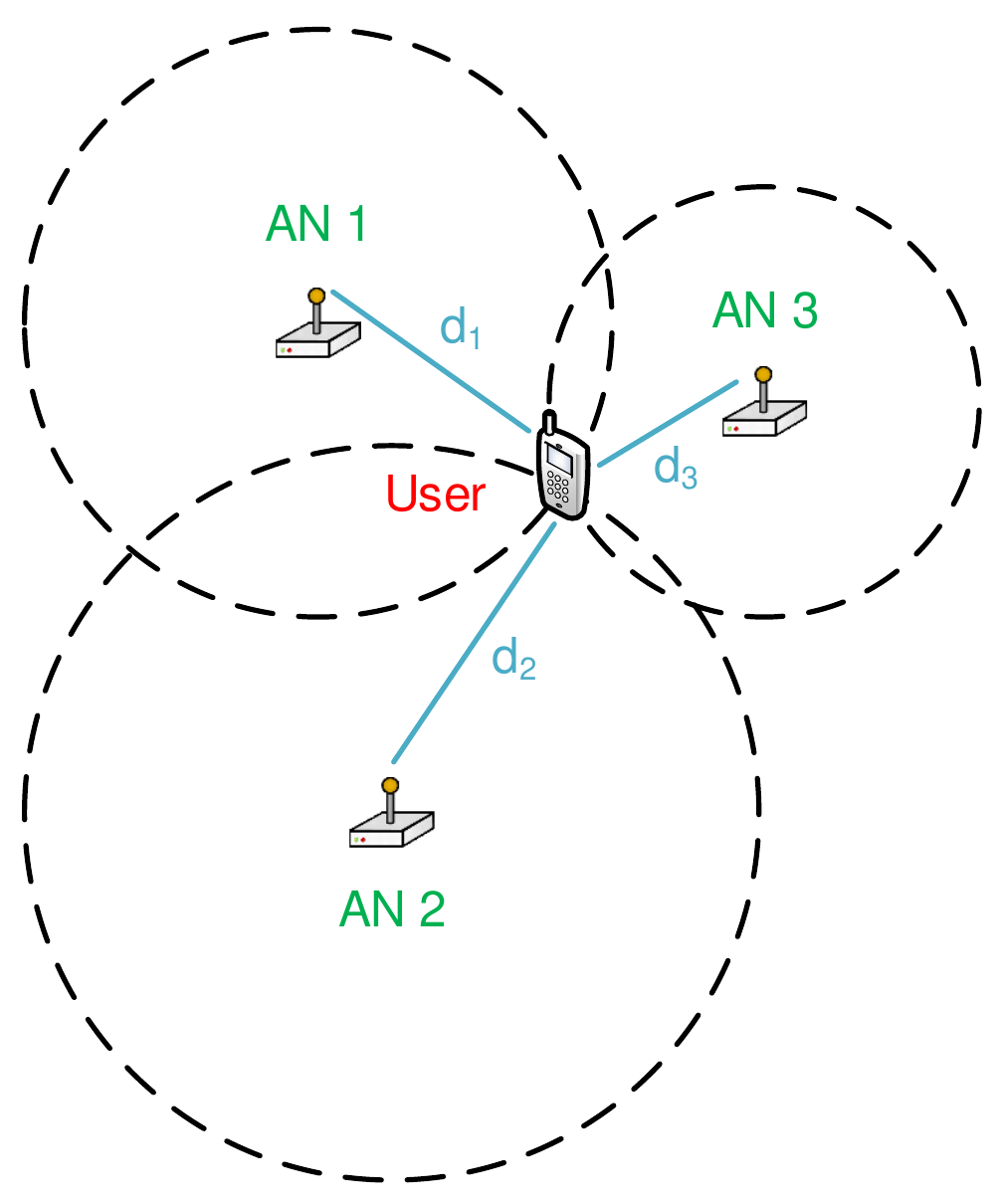}
	\caption{Illustration of localization using wireless signals.}
	\label{localization}
\end{figure}

The localization method is to extract location related information, such as distance or arrival angle, from received signals \cite{RR-2011}. An example of a RF localization system is shown in Fig. \ref{localization}. There are several anchor nodes (ANs) whose locations are known to transmit/receive wireless signals to help locate the user's position. In such a system, the user will acquire the distances to these ANs from received signals and derive its position accordingly. In order to obtain an unique position, it requires at least three ANs. In the sensing systems, objects are located between the Tx and the Rx so that the Rx can detect the changes caused by the objects. Differently, the Rx in a localization system is held by the user since it is the user who wants to know its position. 

\subsection{Measurement Metrics}
To measure the changes or extract location related information, there are various metrics corresponding to different signal properties. The commonly used metrics are reviewed below:

\subsubsection{Received Signal Strength (RSS)} 
RSS indicates how the wireless channel influences the amplitude of wireless signals on average, which is an easy to be acquired metric. In general, the RSS at the Rx with the distance to the Tx being $d$ can be derived through the Log-normal distance path loss model \cite{ST-1992}:
\begin{equation}
	P(d) =  P_T - 10 \alpha \log d + e_{RSS},
\end{equation}
where $P(d)$ is the received RSS measured in Decibel (dB), $P_T$ is the transmitted energy, $\alpha$ is the path loss exponent, and $e_{RSS}$ is the noise for the RSS measurement. For sensing applications, the existence of objects within the sensing area will cause significant signal attenuation, i.e., $\alpha$ will be different, which leads to the variation of RSS measurements. For localization applications, the distance between the user can be obtained from the RSS. 

\subsubsection{Channel State Information (CSI)}
CSI captures the frequency response of the wireless channel. CSI can be obtained by comparing the known transmitted signals in the packet preamble or pilot carriers to the received signals. Different from RSS that only provides the amplitude information, CSI consists of a set of a complex values including both amplitude and phase information, which correspond to multiple orthogonal frequency-division multiplexing (OFDM) subcarriers. Therefore, CSI allows fine-grained channel estimation, and can be mathematically expressed as
\begin{equation}
	\bm{h} = [h_1, \ldots, h_N]^{T},
\end{equation}
where $N$ is the number of subcarrier, and $h_i = a_ie^{j\psi_i}$. Here, $a_i$ is the amplitude of the CSI obtained at the $i$-th subcarriers and $\psi_i$ is the corresponding phase. In general, CSI is more commonly used in sensing applications since it can provide more information to capture movements of objects \cite{YJYMJH-2014}.

\subsubsection{Time of Flight (ToF)}
\begin{figure}[!t]
	\centering
	\includegraphics[width=2.8in]{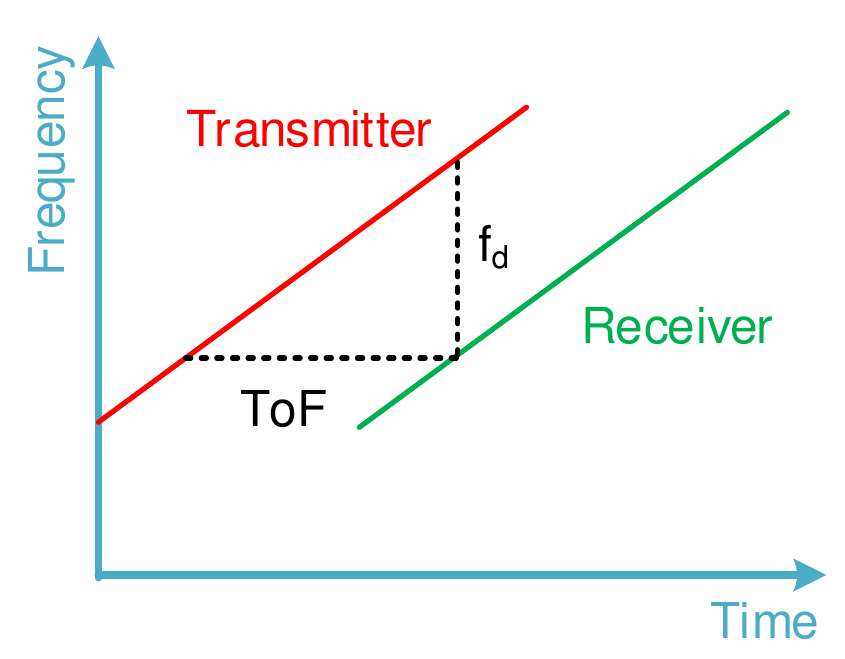}
	\caption{The working principle of FMCW.}
	\label{FWCW}
\end{figure}

ToF is the time span between time instants when a signal is transmitted and received. With the ToF measurement, we can obtain the distance $d$ between the Rx and the Tx as \cite{IC-2009}
\begin{equation}
	c \cdot \tau_{ToF} = d + e_{ToF},
\end{equation}
where $c$ is the speed of the light, $\tau_{ToF}$ is the derived ToF for the transmitted signals, and $e_{ToF}$ is the noise for the ToF measurement. In general, the measurement of ToF can be categorized into the following two types:

\begin{itemize}
\item \textbf{Direct Measurements using OFDM Signals:} Since OFDM is the most commonly used radio waveform, it is important to introduce how to obtain the ToF using OFDM symbols. The received discrete-time OFDM symbol after removing the cyclic prefix (CP) can be expressed as \cite{ZYR-2013}
\begin{equation}
	y_n = e^{\frac{j2\pi\rho n}{N}} \sum\limits_{l = 1}^L h_lx_{lT_s - \tau_l} + \omega_n,
\end{equation} 
where $n$ denotes the index of subcarriers, $x_n$ is the time-domain OFDM waveform sampled at the rate of $1/T_s$, where $T_s$ equals to the symbol duration divided by the number of subcarriers $N$, $h_l$ is the channel coefficient of the $l$-th path with delay $\tau_l$ and $L$ is the total number of paths. $\rho$ is the residual carrier-frequency-offset after frequency synchronization and $\xi_n$ is the additive white Gaussian noise. Assume that $\tau_1 \leq \ldots \leq \tau_L$, and then $\tau_1$ is the ToF that we defined before. Existing methods to estimate the ToF are mainly the MUSIC algorithm~\cite{XK-2004}, the SAGE algorithm~\cite{BMRDK-1999}, and MPLR algorithm~\cite{ZYR-2013}. We take the MPLR algorithm as an example here. The basic idea of the MPLR algorithm is to maximize the peak-to-leaking ratio of estimated channel impulse response, which is a function of the delay. As a result, we can obtain the ToF through the optimization results.

\item \textbf{Frequency Modulated Continuous Wave (FMCW) Radars:} FMCW radars \cite{CS-2017} are the other method to obtain the ToF. It transmits continuous waves instead which allows transmitted signals to stay in a constant power envelop, thus reducing the power and the cost for signal processing. The working principle of FMCW is shown in Fig. \ref{FWCW}. The Tx sends a chirp with linearly increasing frequency, and the Rx will compare the received and transmitted signals at any time instant to figure out the frequency difference $f_d$. Given the slope of the linear chirp $\eta$, the ToF can be directly obtained as
$\tau_{ToF} = \frac{f_d}{\eta}$.
Compared to the direct measurement method, FWCW radars do not require wide bandwidth and high sampling rate, which is more suitable for the systems having a relative narrow bandwidth.
\end{itemize}

\subsubsection{Doppler Shift}
Doppler shift is a property of wireless signals caused by relative movements, which can be utilized to capture movements of sensing targets or positioning users over the observed frequency \cite{A-2005}. For example, in sensing applications, a positive Doppler shift implies that the sensing target moves towards the Rx, while a negative value indicates that the target departs from the Rx. Assume that the sensing target moves at speed $v$ along direction $\phi$ towards the Rx, the resulted Doppler shift can be expressed as 
\begin{equation}
	\Delta f = \frac{2v\cos(\phi)}{c}f,
\end{equation}
where $f$ is the center frequency of the transmitted signals.

\subsubsection{Angle of Arrival (AoA)} 
\begin{figure}[!t]
	\centering
	\includegraphics[width=2.8in]{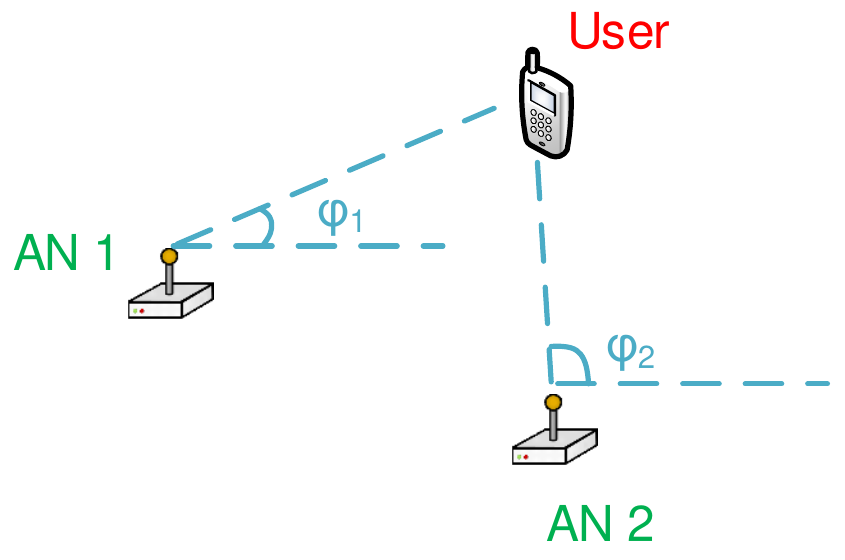}
	\caption{AoA based localization method.}
	\label{AoA}
\end{figure}

\begin{figure*}[t]
	\begin{center}
		\includegraphics[width=0.90\textwidth]{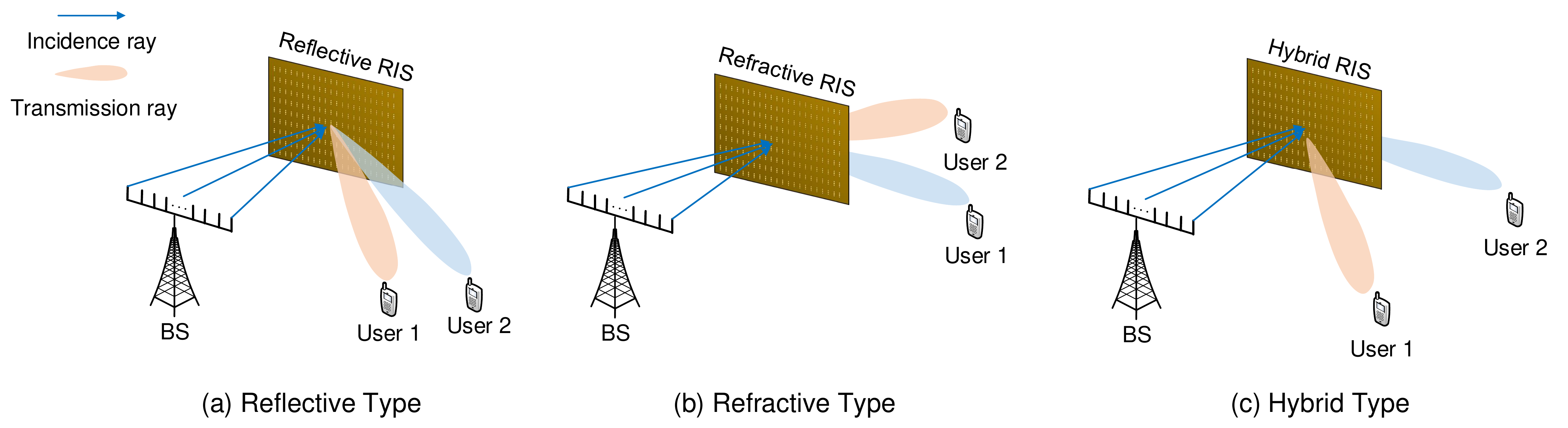}
	\end{center}
	\caption{Different types of RISs: (a) Reflective type; (b) Refractive type; (c) Hybrid type.} \label{type}
\end{figure*}

Sensing and localization can also be facilitated by the detection of AoA, which is also referred to as the direction of arrival (DoA). In sensing applications, by tracking the AoA, the Rx can tell whether the signals are reflected from the sensing area. In localization applications, the AoA-based localization method as shown in Fig. \ref{AoA} only requires two ANs, while the distance-based localization method as shown in Fig. \ref{localization} requires three ANs.

Assume that the locations of the user and the $i$-th AN as $\bm{u} = [u_x,u_y]^T$ and $\bm{a}_i = [a_x^i, a_y^i]^T$, respectively. Therefore, the measured AoA can be written as \cite{RR-2011}
\begin{equation}\label{AoA-location}
	\varphi_i = \arctan\left(\frac{u_y - a_y^i}{u_x - a_x^i}\right) + e_{AoA}^i,
\end{equation}
where $e_{AoA}^i$ is the measurement error. Therefore, we can infer the location of the user through the measured AoAs using (\ref{AoA-location}). To measure the AOAs, the receiver should equip with antenna arrays or directional antennas with spatial resolution capabilities. 

\subsection{RIS Basics}
\label{sub:RIS}

RISs are thin layers of electromagnetic meta-materials capable of shaping radio waves that impinge upon them in ways that the wireless environment can be customized to fulfill specific system requirements \cite{MMDAMCVGJHJAGM-2019}. According to its implementation, the RIS can be categorized into three types~\cite{SHBYZHL-2021}: 

\begin{itemize}
	\item \textbf{Reflective Type:} In this type, the RIS only reflects incident signals towards the users on the same side of the base station (BS), as illustrated in Fig. \ref{type}(a). In the literature, this type of the RIS is also referred to as intelligent reflecting surfaces (IRSs)~\cite{XDYDR-2020}. 
	
	\item \textbf{Refractive Type:} In this type, incident signals will penetrate the RIS and be refracted towards users on the opposite side of the BS as shown in Fig. \ref{type}(b). 
	
	\item \textbf{Hybrid Type:} This type of RIS enables the dual function of reflection and refraction~\cite{HSBYMMLZH-2021}. In other words, the incident signals will be split into two parts: one part is refracted and the other is reflected, as demonstrated in Fig. \ref{type}(c). This type of RIS is also referred to as intelligent omni-surfaces (IOSs) \cite{SHBYZL-2020}.
\end{itemize}

\begin{figure}[t]
	\begin{center}
		\includegraphics[width=2.5 in]{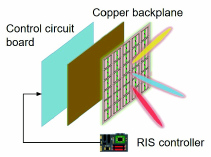}
	\end{center}
	\caption{Components of a reflective RIS.} \label{component}
\end{figure}

Although the RIS is composed of multiple layers, each layer might vary for different types. As illustrated in Fig. \ref{component}, we will take the reflective RIS as an example to show how the RIS is built. An reflective RIS consists of the following three layers:
\begin{itemize}
	\item \textbf{Outer layer} is a two dimensional (2D)-array of RIS elements, which can directly interact with incident signals;
	
	\item \textbf{Middle layer} is a copper plate that can prevent the signal energy leakage;
	
	\item \textbf{Inner layer} is a printed circuit connecting to the RIS controller, which can control the phase shifts of the RIS elements.
\end{itemize}

\begin{figure}[t]
	\begin{center}
		\includegraphics[width=2.5in]{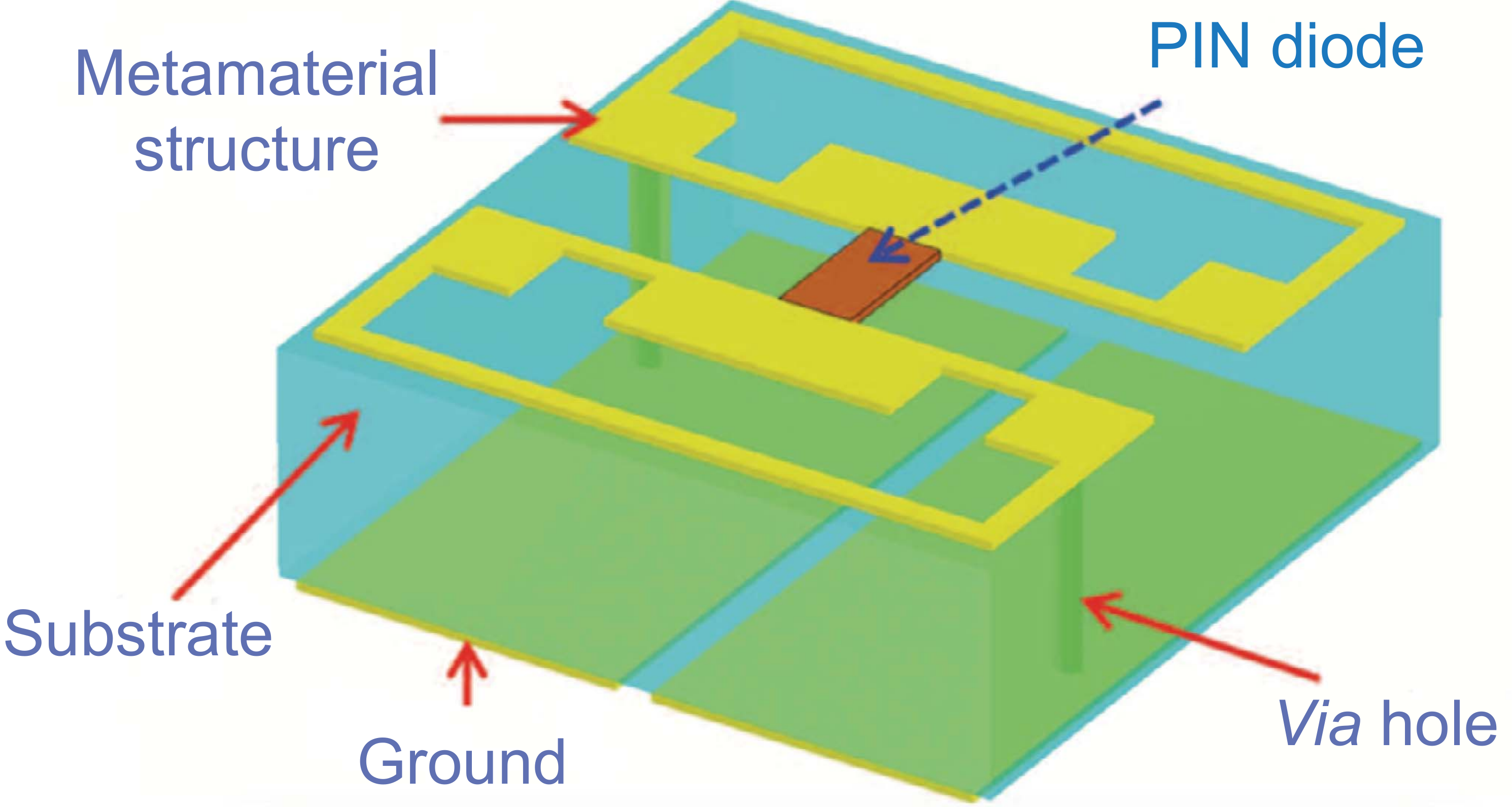}
	\end{center}
	\caption{An example of a programmable meta-material particle.} \label{sec2:element}
\end{figure}

Each RIS element is a low-cost sub-wavelength programmable meta-material particle, whose working frequency can vary from sub-6 GHz to THz~\cite{HWJAAR-2006}. When an EM wave impinges into the RIS element, a current will be induced by the EM wave, and this induced current will emit another EM radiation based on the permittivity $\epsilon$ and the permeability $\mu$ of the RIS. This is how the RIS element controls the wireless signals. An example of the meta-material particle is given in Fig. \ref{sec2:element}. As illustrated in this figure, positive intrinsic negative~(PIN) diodes are embedded in each element. By controlling the biasing voltage through the via hole, the PIN diode can be switched between ``ON" and ``OFF" states. The ``ON" and ``OFF" states of the PIN diodes lead to different values of $\epsilon$ and $\mu$. As a result, this element will have different response to incident signals by imposing different phase shifts and amplitude \cite{BHLLYZ-2020}.

Without loss of generality, we first consider the response of one RIS element. We define the additional phase shift introduced by the RIS as $\theta$. The value of $\theta$ can be continuous if the RIS is incorporated with varactors, while $\theta$ has finite values if it is implemented by PIN diodes. Assume $B$ PIN diodes are implemented in the RIS, we have $K$ possible phase shifts with $K \leq 2^B$, which can be expressed by $\mathcal{K} = \{0, \ldots, 2n\pi/K,\ldots 2(K-1)\pi/K\}, 1\leq n \leq K - 1$~\cite{BHLYZH-2020}. Since the RIS can be divided into three types, we will show how these three types of RISs response to the incident signals in the following.

\begin{figure*}[!t] 
	\center{\includegraphics[width=0.8\textwidth]{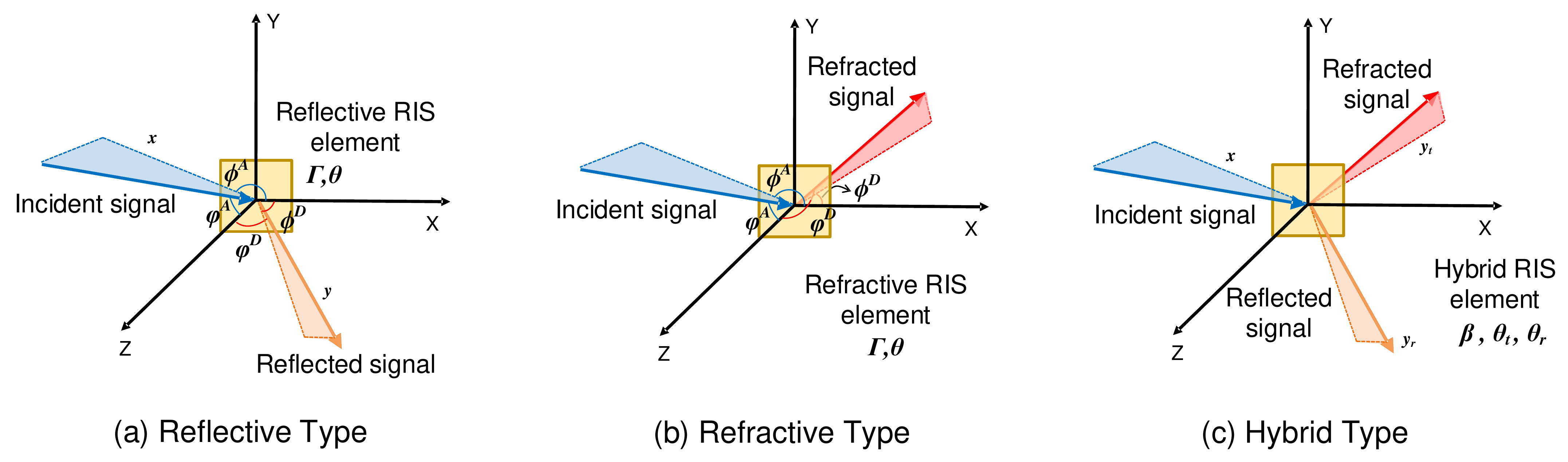}}
	%	\vspace{-1em}
	\caption{Response for an RIS element of three types.}
	\label{response}
\end{figure*}

\subsubsection{Reflective Type}
According to \cite{YXXTJMN-2021}, the response of the RIS element can be written as
\begin{equation}\label{equ:response}
	\gamma = \Gamma e^{-j\theta},
\end{equation} 
where $j$ is the imaginary unit, i.e., $j^2 = -1$. Here, $\Gamma \in [0,1]$ is the amplitude of the RIS response where $\Gamma = 1$ indicates that the incident signals are fully reflected while $\Gamma = 0$ implies that the incident signals are fully absorbed. The response depends on the tuning impedance of the equivalent circuit for each element and mutual impedances (if mutual coupling cannot be ignored) at the ports of the RIS, which generally is influenced by azimuth and elevation angles for incident and reflected signals, i.e., $\phi^A$, $\psi^A$, $\phi^D$, and $\psi^D$, as shown in Fig. \ref{response}(a). Moreover, $\Gamma$ and $\theta$ are usually sensitive to the working frequency. The reflection coefficients will vary when the same RIS receives signals with different frequency. However, the RIS will be designed to operate over a predefined band where the phase shift and amplitude can be regarded to be unchanged over the considered bandwidth.   

\subsubsection{Refractive Type}
As shown in Fig. \ref{response}(b), similar to the reflective type, the response of a refractive RIS element can also be expressed in the form of (\ref{equ:response}). The only difference is that the incident signals fully penetrate the RIS element when $\Gamma = 1$.

\subsubsection{Hybrid Type}
The hybrid RIS element has the functions of both reflection and refraction. Therefore, the RIS will first split the energy of incident signals into two parts: one for refractive signals and the other for reflective signals. To quantify the energy separation, we introduce a metric $\beta \in [0, +\infty)$, which is the power ratio of reflected signals to transmitted signals \cite{SHBYMZHL-2021}. Therefore, we assume that the no energy leakage, and the response of a hybrid RIS element to reflected and transmitted signals can also be expressed in the form of (\ref{equ:response}), where the phase shifts to reflected and transmitted signals might be different. Therefore, the reflective and refractive responses can be expressed as
\begin{align}
	\gamma_{refl} &= \sqrt{\frac{\beta}{1+\beta}}\Gamma_{refl} e^{-j\theta_{refl}},\\
	\gamma_{refr} &= \sqrt{\frac{1}{1+\beta}}\Gamma_{refr} e^{-j\theta_{refr}},
\end{align} 
where $\Gamma_{refl}$ and $\Gamma_{refr}$ are the amplitudes for reflection and refraction, and $\theta_{refl}$ and $\theta_{refr}$ are the phase shifts for reflection and refraction, respectively. It is worthwhile to point out that the hybrid type will be reduced to the reflective type with $\beta = +\infty$, and the refractive type with $\beta = 0$.

In the following of this paper, we take the reflective type RISs as examples, and the term ``RIS" typically refers to the reflective RIS for brevity. The same idea can also be applied to refractive or hybrid RISs.

\subsection{RF Signal Modeling with RISs}
\label{sub:RF signal}

\begin{figure}[!t] 
	\center{\includegraphics[width=2.8in]{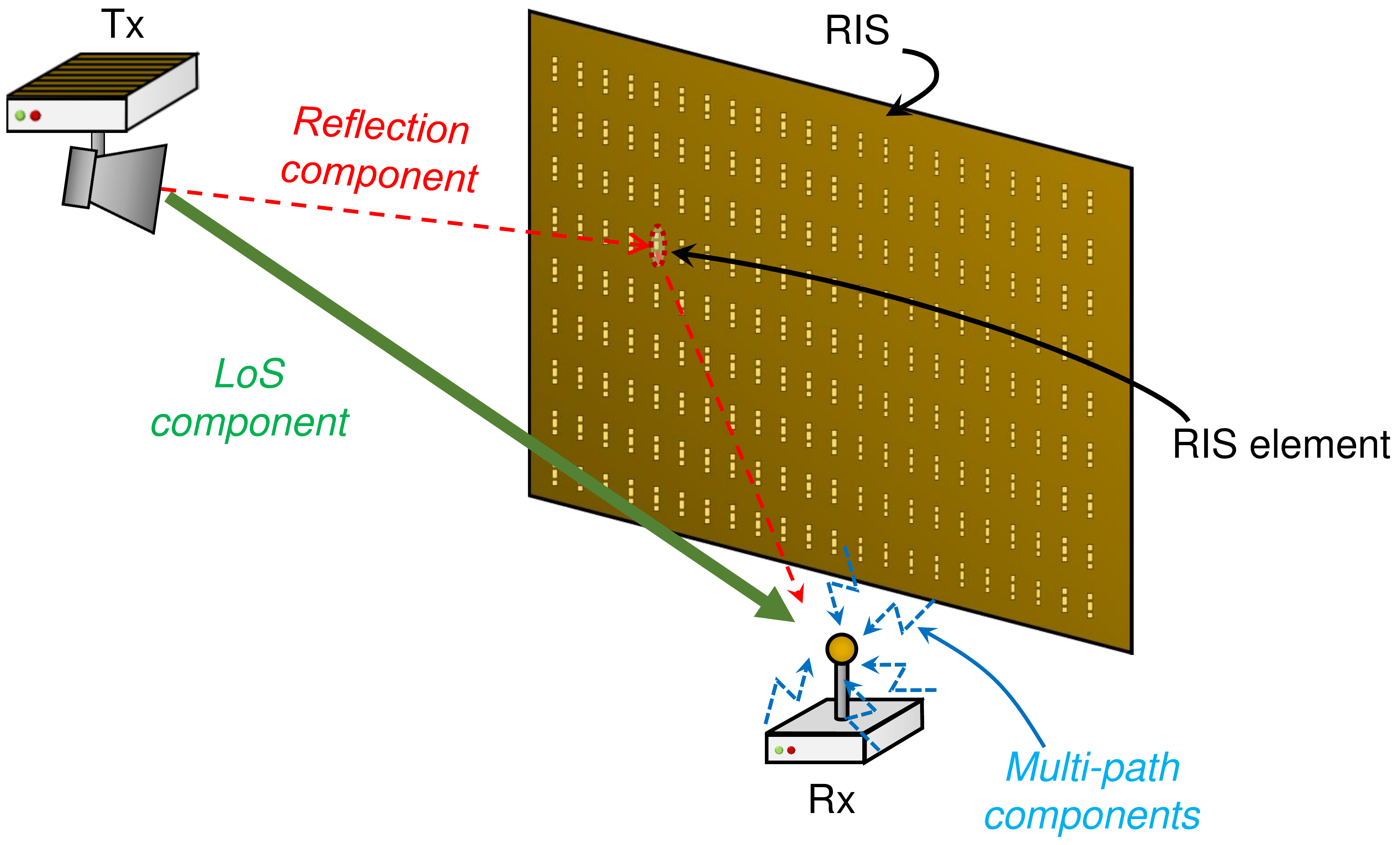}}
	%	\vspace{-1em}
	\caption{A single-user communication system with an RIS.}
	\label{channel}
\end{figure}

With an RIS, a single-user communication system is shown in Fig. \ref{channel}. In the following, we will introduce how to model the signals in an RIS-aided communication system. We will start with a single subcarrier, and then extend the modeling to an OFDM system.

\subsubsection{Single Subcarrier}
Over the $n$-th subcarrier, the received signals are composed of three components, i.e., LoS component,  reflection component, and multi-path component, as elaborated below:

\textbf{LoS Component} 
The LoS component indicates the direct signal path from the Tx to the Rx. Denote $h_{los}$ as the channel gain for the LoS component. Based on~\cite{A-2005}, $h_{los}$ can be expressed as
\begin{equation}
	\label{LoS_Channel}
	h^{n}_{los} = \frac{\lambda}{4\pi} \cdot \frac{\sqrt{g_{T}g_{R}}\cdot e^{-j2\pi d_{los}/\lambda_n}}{d_{los}},
\end{equation}
where $\lambda_n$ is the wavelength of the signals transmitted on the $n$-th subcarrier, $g_{T}$ and $g_{R}$ denote antenna gains of the Tx and the Rx, respectively, and $d_{los}$ is the distance between the Tx and the Rx.
	
\textbf{Reflection Component} 
The reflection component are the LoS paths from the Tx to the Rx via the reflections of the RIS, where each RIS element corresponds to one reflection path. Assume that the RIS consists of $M$ elements and define $h_{m}$ as the gain of the reflection path via the $m$-th RIS element. Based on~\cite{HBZHL-2021,WMXJYMYSQT-2021}, $h_{m}$ can be written as 
\begin{equation}
	\label{RIS_Channel}
	h^{n}_{m}= \frac{\lambda \sqrt{g_{T}g_{R}} \gamma_m  e^{-j2\pi (d^T_m+d^R_{m})/\lambda_n}}{8\pi^{3/2} d^T_{m} d^R_{m}},
\end{equation}
where $\gamma_m$ is the response of the $m$-th RIS element as defined in (\ref{equ:response}), $d^T_m$ and $d^R_m$ denotes the distances from the $m$-th RIS element to the Tx and the Rx, respectively. Note that the multi-path component will also reflected by the RIS. However, the channel gain of the reflection of the multi-path component is much less than that of the LoS reflection component, and thus, the reflection of the multi-path component is neglected here.
	
\textbf{Multi-path Component}
The environmental scattering paths account for the signals paths between the Tx and the Rx, which involve complex scattering from surrounding environment. We denote $h^n_{sc}\in\mathbb C$ as the equivalent gain of all the environmental scattering paths.

Based on above notations, the received signals over the $n$-th subcarrier can be expressed as
\begin{equation}
	\label{RSS}
	y_n= \left( h^n_{los} + \sum_{m=1}^M h^n_{m} + h^n_{sc}\right) x_n + \omega_n,
\end{equation}
where $x_n$ denotes the transmitted symbol over the $n$-th subcarrier, and $\omega_n$ is the noise where $\omega_n \sim\mathcal{CN}(0,\sigma^2)$ with $\sigma^2$ being the noise power. 

\subsubsection{OFDM Systems}
Let $\bm{x} = [x_1, \ldots, x_N]^T$ be the OFDM symbol. The OFDM symbol is first transformed into the time domain via an $N$-point inverse discrete Frourier transform (IDFT) and then is appended by a CP. At the receiver, after removing the CP and performing $N$-point Frourier transform (DFT), the equivalent baseband signals in the frequency domain at the receiver can be expressed as
\begin{equation}	
	\label{equ: received signal}
	\bm{y} = \bm{X}(\bm{h}_{los} + \bm{h}_{reflect} + \bm{h}_{sc}) + \bm{\omega},
\end{equation}  
where $\bm{y} = [y_1, \ldots, y_N]^T$ is the received OFDM symbol, $\bm{X} = \text{diag}(\bm{x})$ is the diagonal matrix of the OFDM symbol $\bm{x}$, $\bm{h}_{los} = [h_{los}^1, \ldots, h_{los}^N]^T$ is the channel response of the LoS component, $\bm{h}_{reflect} = \left[\sum_{m=1}^M h^1_{m}, \ldots, \sum_{m=1}^M h^N_{m}\right]^T$ is the channel response of the reflection component, $\bm{h}_{sc} = [h_{sc}^1, \ldots, h_{sc}^N]^T$ is the channel response of the multi-path component, and $\bm{\omega} = [\omega_1, \ldots, \omega_N]^T$ is the noise. According to (\ref{equ: received signal}), we can adjust the phase shifts of the RIS to customize the received signals for a certain system requirement.

\section{RIS-aided Sensing}
\label{Sensing}
In this section, we will introduce the RIS-aided RF sensing applications. In traditional RF sensing problems, ones need to optimize the decision function to map the received signals to sensing results. However, the use of the RIS introduces two unique challenges in system design. \emph{First}, RIS configurations need to be carefully designed to provide favorable wireless propagation environment for sensing applications. Different from the design for communications, which aims to maximize the received signal-to-noise ratio~(SNR), the RIS configuration design for sensing applications is to enhance the differences of signals in the presence of different sensing targets, which makes the receiver easier to distinguish these targets. \emph{Second}, as the RIS can manipulate the received signals, the decision function design is highly coupled with the RIS configurations, which makes the design of the decision function challenging.

In general, the RF sensing techniques can be divided into two types according to the used RF signals. One is to utilize commodity signals, such as WiFi and cellular signals, which we refer to as \emph{MetaSensing}. The other one is to use customized signals, i.e., radar, which is referred to as \emph{MetaRadar}. Different from the first type, the radar sensing can adjust the transmitted waveform, and thus can have a better sensing accuracy. However, it requires a pair of extra transceiver for signal transmission and reception, which is more likely to be used in the applications sensitive to the accuracy, such as autonomous driving. In Section \ref{Subsec:sensing} and Section \ref{Subsec:radar}, we will present the details about MetaSensing and MetaRadar systems to address the above challenges, respectively. 

\subsection{MetaSensing: Sensing with Commodity Signals}
\label{Subsec:sensing}
\begin{figure}[!t]
	\center{\includegraphics[width=2.8in]{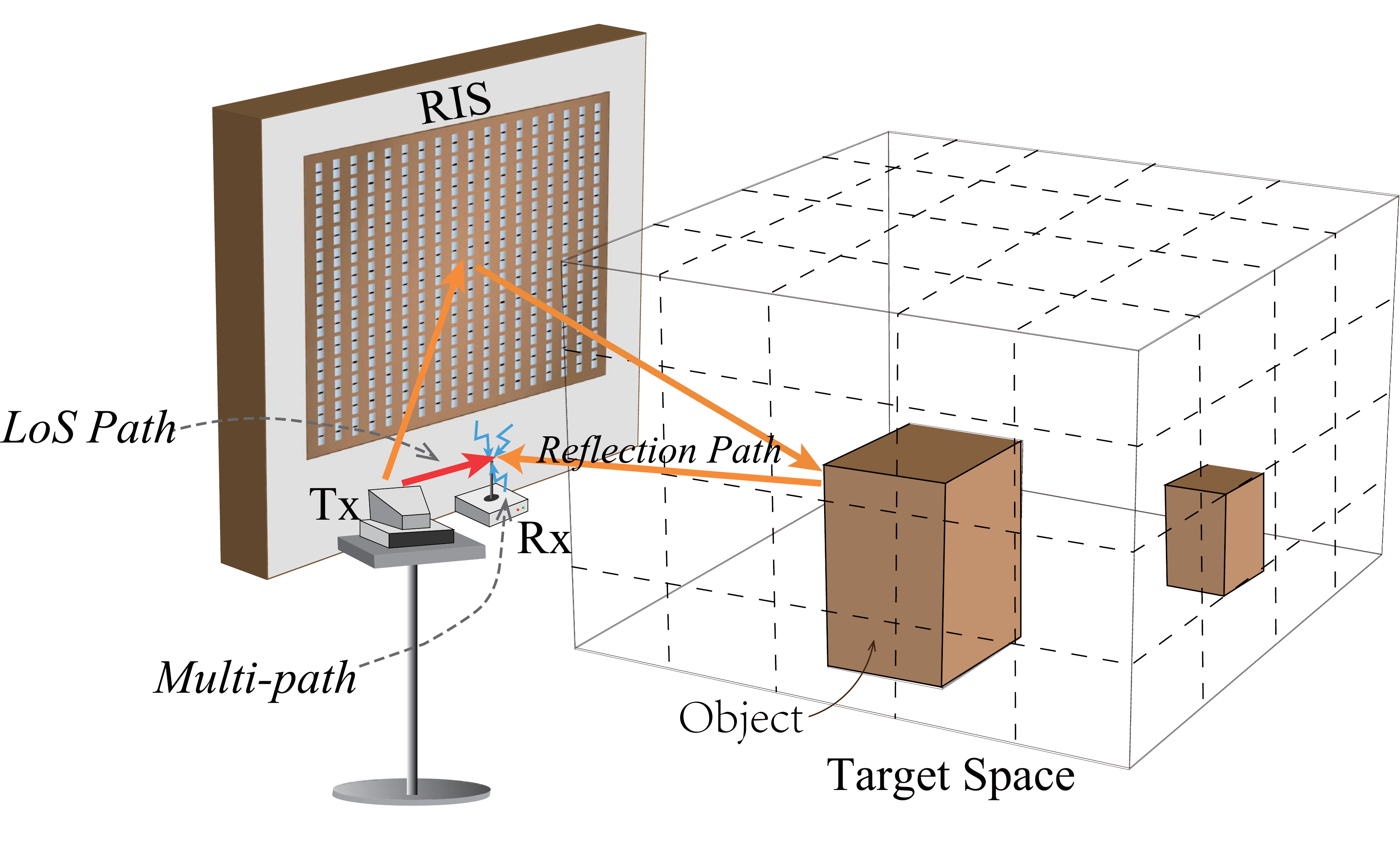}}
	\caption{Illustration for MetaSensing systems.}
	\label{sensing:seytem}
\end{figure}

A general MetaSensing system with commodity signals is illustrated in Fig.~\ref{sensing:seytem}. In such a system, there exist a pair of the Tx and Rx, an RIS, and a target space where the objects (or human bodies) are located. Here, the Tx and Rx are commercial devices, for example, Wi-Fi access points and smartphones. The target space is a cubical region that is discretized into $Q$ uniform \emph{space blocks}. The transmitted signals are customized by the RIS before entering into the target space. The customized signals are further reflected by the objects in the target space and received by the Rx unit. As a result, the Rx can map the received signals to the sensing results, where the received signals include the LoS and reflected links.

\begin{figure}[!t] 
	\center{\includegraphics[width=3.0in]{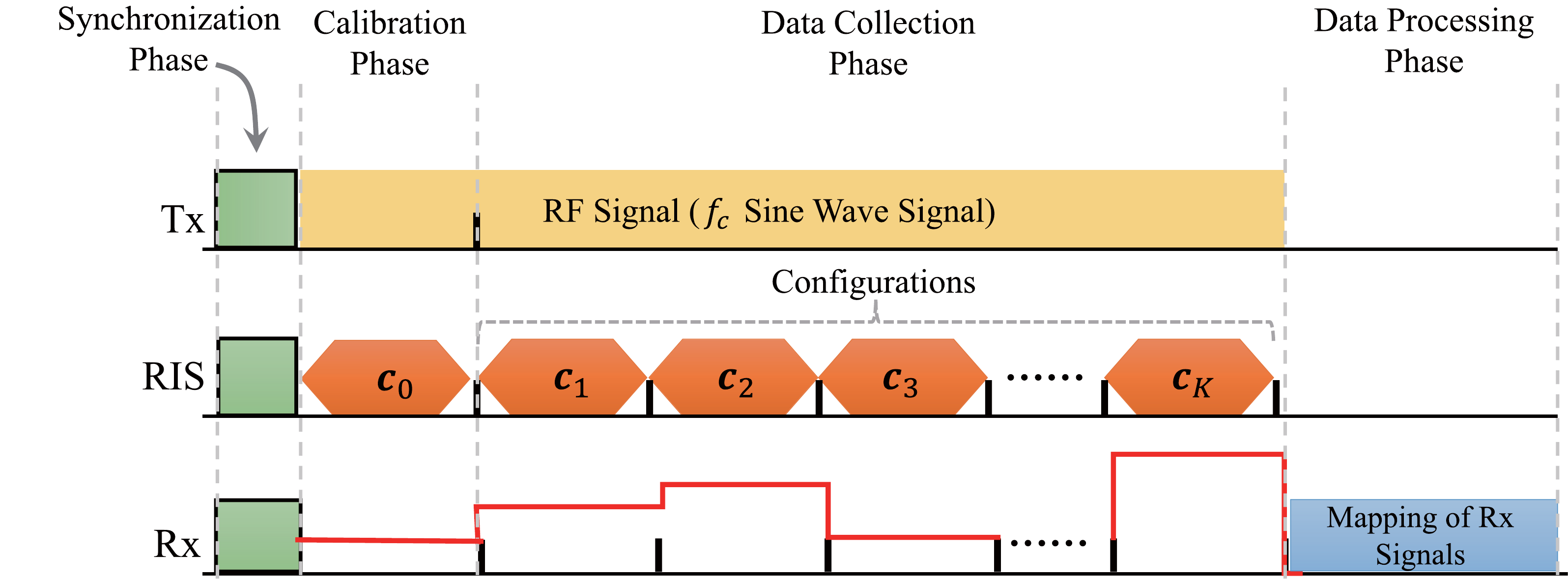}}
	\caption{A cycle of the sensing protocol.}
	\label{sensing: protocol}
\end{figure}

To synchronize the RIS, the Tx, and the Rx, a \emph{sensing protocol} is proposed \cite{JHKMZL-2021}. In the protocol, the timeline is divided into \emph{cycles}, where the Tx, the Rx, and the RIS are operated in a synchronized and periodic manner. As shown in Fig.~\ref{sensing: protocol}, each cycle consists of four phases: 
\begin{itemize}
\item \textbf{Synchronization Phase:} the Tx transmits a synchronization signal to the RIS and to the Rx, which identifies the start time of a cycle.

\item \textbf{Calibration Phase:} As the received LoS path contains no information of the target space, we generate a reference signal in this phase which will be used to subtract the LoS path. To be specific, the RIS is set to a default configuration and the Rx records the received reference signal $y_0$.

\item \textbf{Data Collection Phase:} The timeline in this phase is equally divided into \emph{frames}. During this phase, the Tx  continuously transmits the RF signal, while the RIS changes its configuration at the end of each frame, as shown in Fig.~\ref{sensing: protocol}. The received signals are denoted by $\bm{y}$. To remove the LoS path, the received signals at these frames are subtracted by the reference signal. Specifically, the differences between the received signals and the reference signals constitute a \emph{measurement vector} $\bm{\hat{y}}$, where
\begin{equation}
	\label{equ2: measurement vector}
	\bm{\hat{y}} = \bm{y} - \bm{y}_0 = \tilde{\bm{\Gamma}} \bm \nu x + \tilde{\bm \omega}.
\end{equation}
Here, $\bm{y}_0$ is a reference vector where all the elements are the reference signal $y_0$ obtained in the previous phase, and $x$ is the transmitted signal. $\bm{\nu}$ is the reflection coefficients of these space blocks, which can be used to determine the existence of target on each block. $\tilde{\bm \omega}$ is the noise difference between the received signals and the reference signal, which is still Gaussian. $\tilde{\bm{\Gamma}}$ is the difference between the channel gains under the RIS configurations selected in this phase and that under the default configuration.

\item \textbf{Data Processing Phase:} The Rx maps the measurement vectors obtained in the data collection phase to the sensing results using a decision function.
\end{itemize}

To improve the sensing performance, the configuration of the RIS and the decision function at the Rx need to be optimized. In the following, we will introduce two use cases. The first one is known objects, where at least partial information of the targets is known. The second one is unknown objects, where no prior information about the targets is given.

\subsubsection{Case I A Known Object Set}
For a known object set, we take the posture recognition as an example, where the set of the possible postures $\mathcal{I}$ is known \cite{JHBLLYZH-2020}, and we want to figure out what the posture is from the received signals. To quantify the sensing performance, we define the weighted cost caused by the false recognition under the decision function $L$ as \emph{average false recognition cost}, where
\begin{equation}
	\label{equ: classification cost2}
	\Psi_{L} =\sum_{i, i'\in\mathcal{I},~i\neq i'} p_i \chi_{i,i'} \int \Pr(\bm{\hat{y}}_i|\bm \nu_i) L_{i'}( \bm{\hat{y}}_i )  d\bm{\hat{y}}_i.
\end{equation}
Here, $p_i$ is the prior probability of the $i$-th human posture, $\bm \nu_i$ is the reflection coefficient vector of the $i$-th posture, $\chi_{i,i'}$ is the cost of recognizing the $i$-th posture as the $i'$-th posture, $\Pr(\bm{\hat{y}}_i|\bm \nu_i)$ denotes the probability for the measurement vector to be $\bm{\hat{y}}_i$ given $\bm \nu_i$, and $L_{i'}(\bm{\hat{y}}_i)$ is the probability for the Rx to map the measurement vector $\bm{\hat{y}}_i$ to the $i'$-th posture.

\textbf{RIS Configuration Optimization:} This problem is to minimize the average false recognition cost by optimizing the configuration matrix which is involved in $\tilde{\bm{\Gamma}}$. Moreover, the optimized configuration matrix for specific coefficient vectors may be sensitive to the subtle changes of the postures. Therefore, we will reformulate the objective for a general posture recognition scenario.

Based on the observations that 1) most of the space blocks are empty, and thus have zero reflection coefficients, and 2) for the blocks where the human body lies, only those that contain the surfaces of the human body with specific angles can reflect the incidence signals towards the Rx and have non-zero reflection coefficients, the reflection vector for the target space $\bm \nu$ is sparse and can be reconstructed through compressive sensing. According to the theory in \cite{M-2007}, to minimize the loss of reconstruction for sparse target signals, we can minimize the averaged mutual coherence of $\tilde{\bm{\Gamma}}$, which is defined as
\begin{equation}
	\label{equ: mutual coherence}
	\mu(\tilde{\bm{\Gamma}})= \frac{1}{Q(Q-1)}\sum_{q,q',q\neq q'}\frac{|\tilde{\bm  \gamma}_q^T \tilde{\bm  \gamma}_{q'}|}{\| \tilde{\bm \gamma}_q\|_2 \|\bm \tilde{\bm \gamma}_{q'}\|_2},
\end{equation}
where $ \tilde{\bm \gamma}_q$ is the $q$-th column of $\tilde{\bm \Gamma}$, and $\|\cdot\|_{2}$ denotes the $l_{2}$-norm.

\textbf{Decision Function Optimization:} This problem is to minimize the cost by optimizing the decision function $L$. To solve this problem efficiently, we employ a neural network to approximate the decision function. The input is the measurement vector $\bm{\hat{y}}$ and the output is the probability distribution over all the postures in $\mathcal{I}$. This neural network can be trained by the back-propagation algorithm.

\begin{figure}[!t] 
	\center{\includegraphics[width=3.0in]{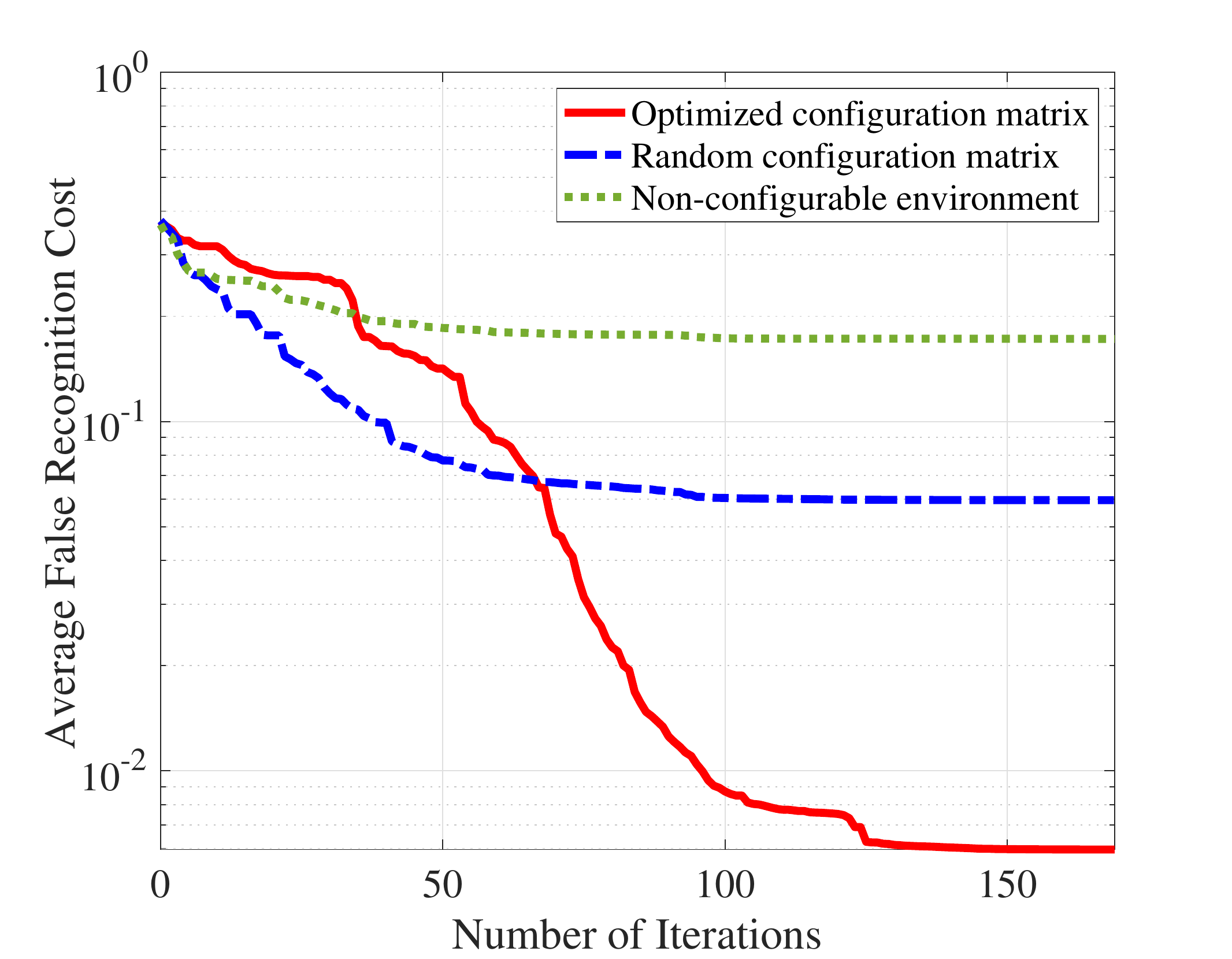}}
	\caption{Average false recognition cost vs. the number of training iterations with 10 cycles.}
	\label{sensing: simulation known}
\end{figure}

Fig.~\ref{sensing: simulation known} shows the average false recognition cost vs. the number of training iterations for the neural network in the decision function optimization. Here, the costs of true and false recognition are set to $0$ and $1$, respectively, i.e., $\chi_{i,i'} = 0$ if $i=i'$, and otherwise, $\chi_{i,i'} = 1$. It can be observed that the converged value of average false recognition cost with the proposed RIS configuration optimization method is less than 10\% of that obtained with a random RIS configuration. Moreover, comparing to the non-reconfigurable environment case, i.e., without the assist of the RIS, we can observe that the capability of the RIS to customize the environment helps the RF sensing system to significantly reduce the average false recognition cost.

\subsubsection{Case II An Unknown Object Set}
\begin{figure*}[!t] 
	\center{\includegraphics[width=0.8\textwidth]{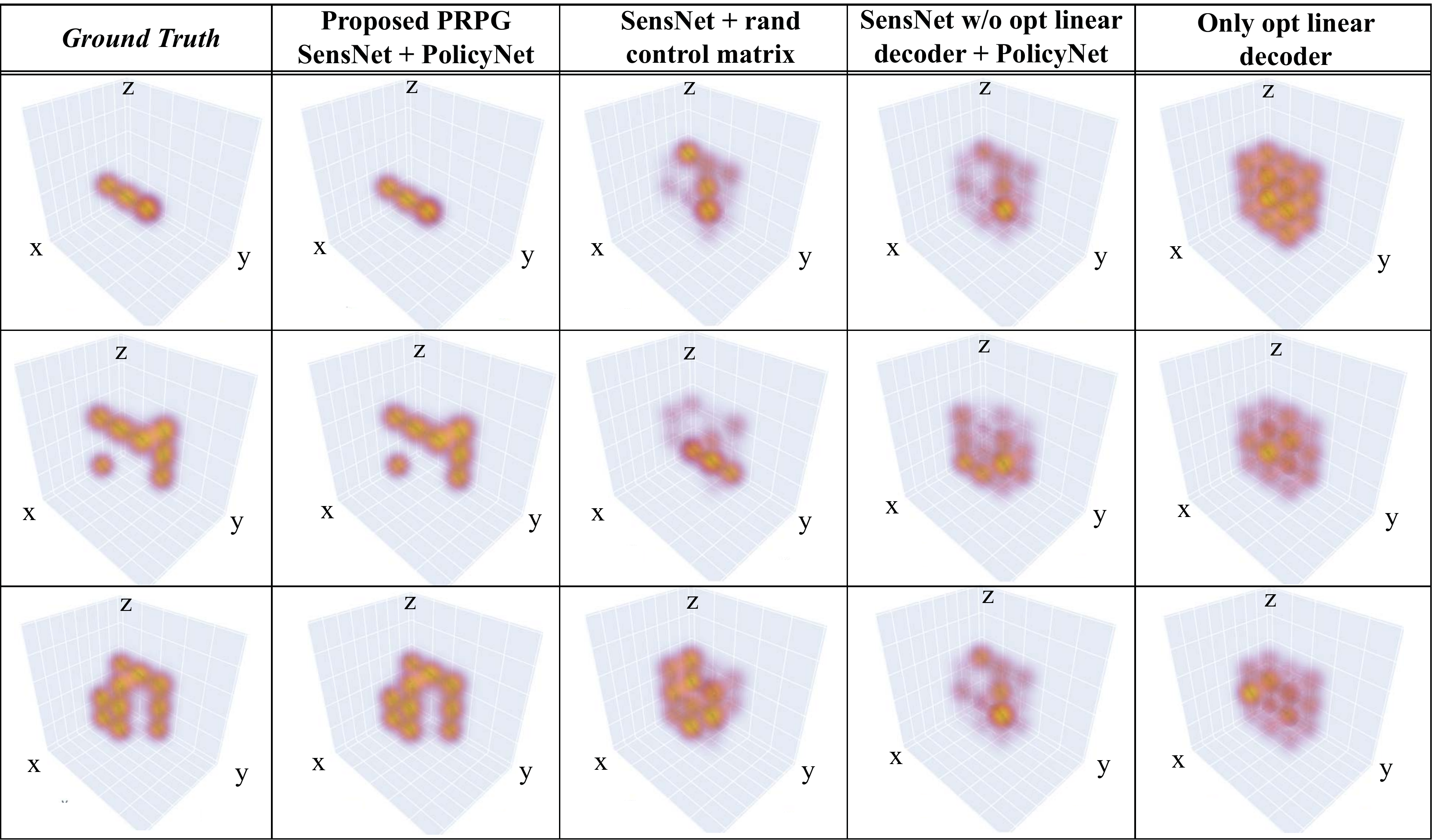}}
	\caption{Sensing result comparisons for different shapes of objects.}
	\label{sensing: simulation unknown}
\end{figure*}

For an unknown object set, we have no information about the number and shape of possible objects, and we aim to obtain a 3D sketch of the objects in a given space from the received signals. As a result, we cannot use the false recognition cost to quantify the sensing accuracy. Instead, we use the \emph{cross-entropy loss} to measure the sensing accuracy \cite{IYAY-2016}. To be specific, the cross-entropy loss is generally used to calculate the difference between the measured probability distribution and the ground-truth distribution over the target space, where
\begin{equation}
	\label{equ2: def of sensing loss}
	\Psi = - \mathbb E_{ \bm \nu} \left[\sum_{q=1}^Q  p_{q}(\bm\nu)  \ln(\hat{p}_{q}) + (1-p_{q}(\bm\nu))  \ln(1-\hat{p}_{q}) 
	\right].  
\end{equation}
Here, $p_q(\bm \nu)$ is a binary variable indicating the object existence in the $q$-th space block. In other words, $p_q(\bm \nu) = 0$ if $|\nu| = 0$, and otherwise $p_q(\bm \nu) = 1$. $\hat{p}_q$ is the estimated results obtained from measurement vector $\bm{\hat{y}}$ using a decision function $f(\bm{\hat{y}})$.

This problem is to minimize the cross-entropy loss by optimizing the configuration matrix which is involved in $\bm{\Gamma}$. As we do not have any information about the objects, the sparsity assumption may not hold in this case. Therefore, the compressive sensing method cannot be applied anymore. As each RIS element has a finite number of phase shifts, the optimization problem can be considered as a decision optimization problem, which can be formulated as a Markov decision process (MDP):
\begin{itemize}
	\item \emph{State}: The state of the environment includes the index of the current frame, the index of the RIS element to adjust its phase shift, and the RIS configuration matrix including the configurations of the RIS over all the frames in the data collection phase. The state is called \emph{terminal state} when the phase shifts of all the RIS elements over all the frames are determined.
	
	\item \emph{Action}: In each state, the RIS element indicated by the index adjusts its phase shift under the current configuration of the RIS.
	
	\item \emph{Transition}: A non-terminal state will transit to the state where the index of RIS element increases by $1$. Moreover, if the last RIS element selects its phase shift in the current state. Then in the next state, the index of the frame will increase by $1$ and the index of the RIS element will be reset to $1$. During the transition, the RIS configuration matrix is updated according to the phase shift adjusted by the RIS element in the current state.
	
	\item \emph{Reward}: The reward is defined as the negative cross-entropy loss of the mapping of the received signals given the configuration determined in the terminal state. If the terminal state has not been reached, the reward for the state transition is set to be zero.
\end{itemize}

Under such an MDP framework, we propose a deep reinforcement learning algorithm to jointly optimize the RIS configuration and decision function. To be specific, the reinforcement learning algorithm consists of two phases as follows, and these two phases proceed iteratively until it converges. Please refers to \cite{JHKMZL-2021} for more details. Based on the results obtained after the convergence, we can further perform semantic recognition and segmentation to obtain meaningful representations of the objects \cite{JHKZHL-2022}.

\textbf{RIS Configuration Optimization Phase:} The RIS starts from an initial state and adopts a policy function to select actions in each state until it reaches the terminal state. As the set of feasible actions is large, we use a neural network to approximate the policy function, called \emph{policy network}, where the input is the states while the output is the probability distribution of the actions. The policy network is trained to maximize the accumulated rewards.

\textbf{Decision Function Optimization Phase:} Similarly, the decision function is also approximated by a neural network, called \emph{sensing network}. The input is the received signals $\hat{\bm{y}}$ under the configurations obtained by the policy network, and the output is the sensing result. The sensing network is trained to minimize the cross entropy defined in (\ref{equ2: def of sensing loss}).

In Fig.~\ref{sensing: simulation unknown}, we show the ground-truths and the sensing results for different shapes of objects. The schemes for these results are 1) ground truths for comparison; 2) the proposed scheme where both sensing and policy networks are used; 3) the sensing network is used while a random RIS configuration is adopted; 4) the policy network is utilized while a model-aided decoder contained in the sensing network is removed; 5) Only the model-aided decoder is used. From the comparison, we can observe that the proposed algorithm outperforms other benchmark algorithms to a large extent, and find that both sensing and policy networks contribute to improving the sensing accuracy. Moreover, we can observe that the proposed algorithm obtains the accurate sensing results despite the different shapes of the objects.

\begin{figure*}[!t]
	\centering
	\subfloat[]{
		%\label{a1} %% label for first subfigure
		\includegraphics[width = 0.4\textwidth]{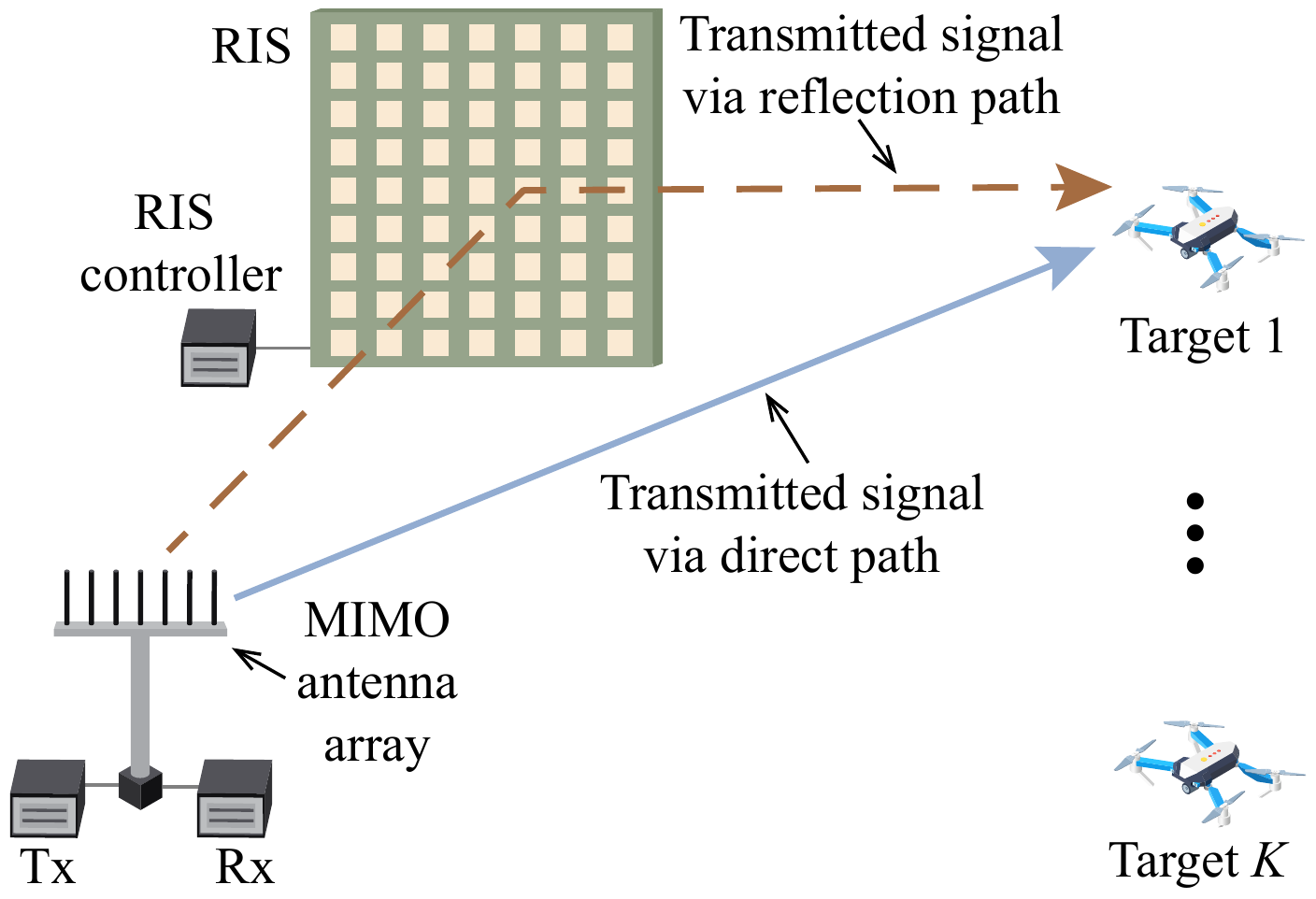}
	}
	\hspace{0.4in}
	%%----start of second subfigure----
	\subfloat[]{
		%\label{a1} %% label for first subfigure
		\includegraphics[width = 0.4\textwidth]{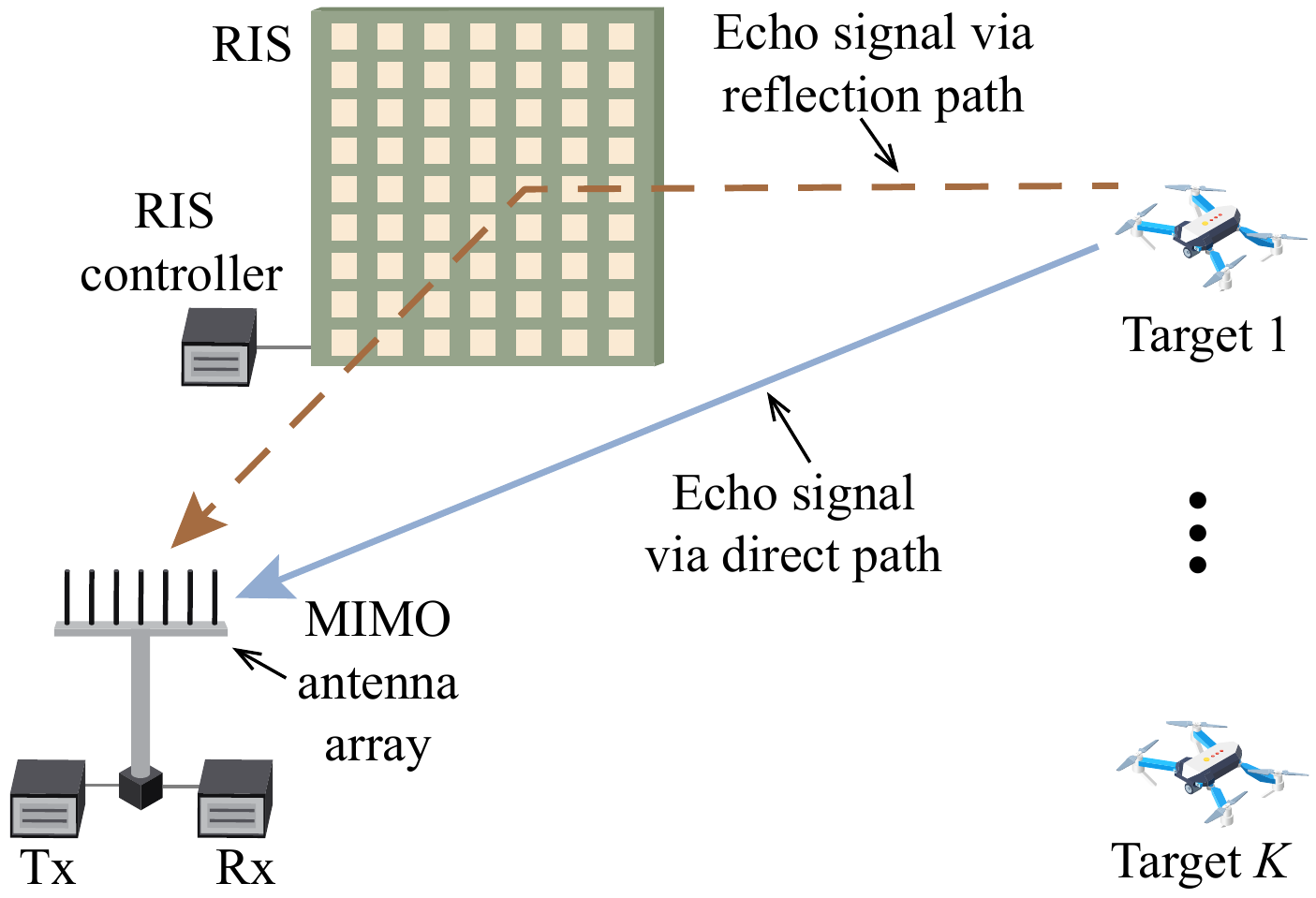}
	}
	\caption{A MetaRadar system: (a) Transmission mode; (b) Reception mode.}
	\label{f_model}
	%\vspace{-6mm}
\end{figure*}

It is worth pointing out that Case I is a special case of Case II. In other words, the methods in Case II can also be used for Case I. However, the complexity of the method proposed for Case I is typically lower than that for Case II since we can use the prior information to reduce the feasible set.

\subsection{MetaRadar: Sensing with Radar Signals}
\label{Subsec:radar}

A multi-target detection scenario using a MetaRadar is illustrated in Fig. \ref{f_model}. The MetaRadar is composed of a Tx, an Rx, a MIMO antenna array connected with the Tx and Rx, and an RIS \cite{HHBKZL-2021}. By deploying an RIS in the radar system, we can improve the overall channel conditions between the antenna array and sensing targets. The MetaRadar has two modes, i.e., transmission and reception modes. In the transmission mode, the Tx first generates signals according to designed waveforms, and then radiates the signals through the MIMO antenna array towards the targets via both direct and reflection paths, as illustrated in Fig.~\ref{f_model}~(a). Then, the MetaRadar system converts to the reception mode, where the antenna array receives the echo signals reflected by the targets. The received signals will be delivered to the Rx in order to detect and locate targets.

\begin{figure*}[!t]
	\centering
	\subfloat[]{
		%\label{a1} %% label for first subfigure
		\includegraphics[width = 0.4\textwidth]{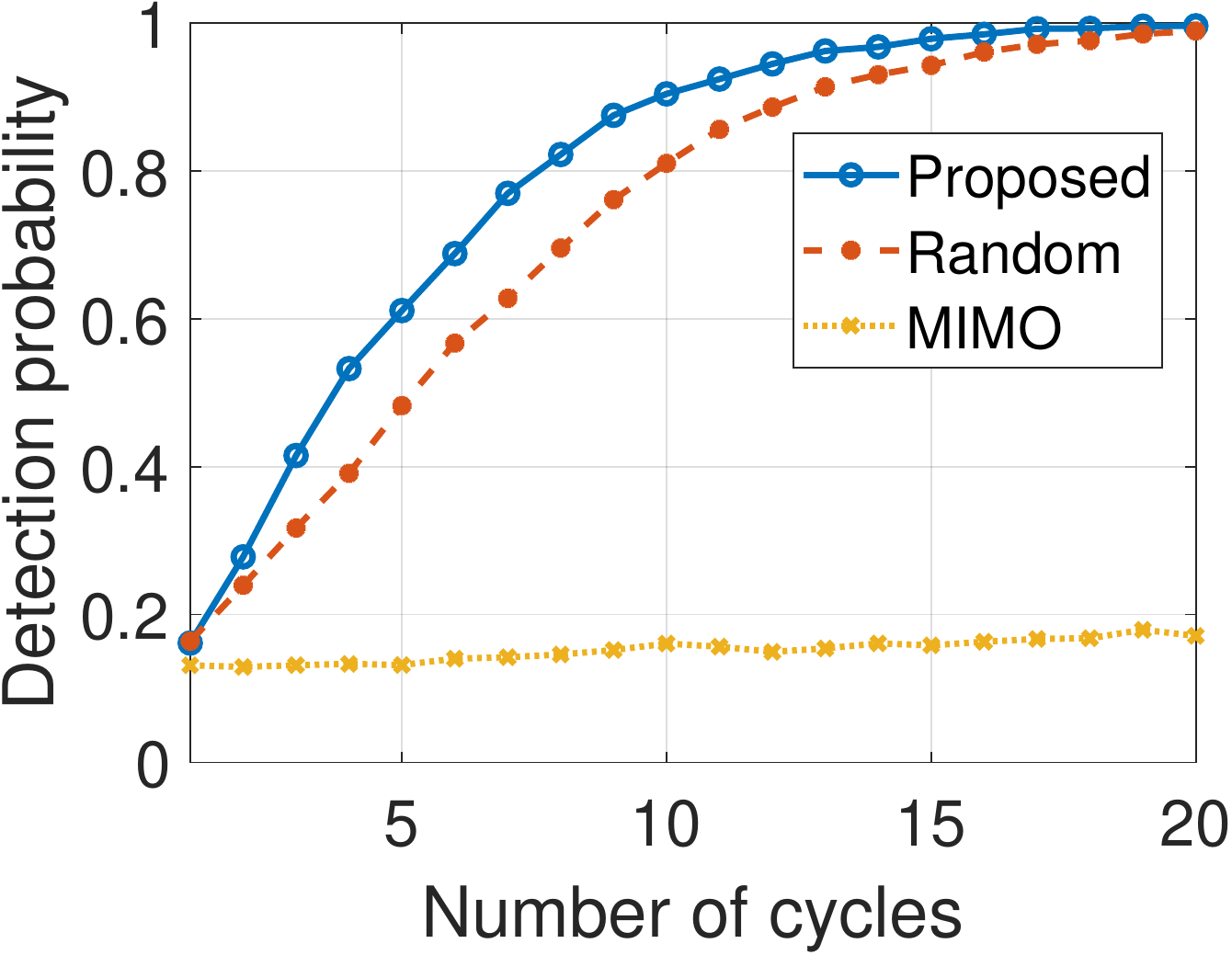}
	}
	\hspace{0.4in}
	%%----start of second subfigure----
	\subfloat[]{
		%\label{a1} %% label for first subfigure
		\includegraphics[width = 0.4\textwidth]{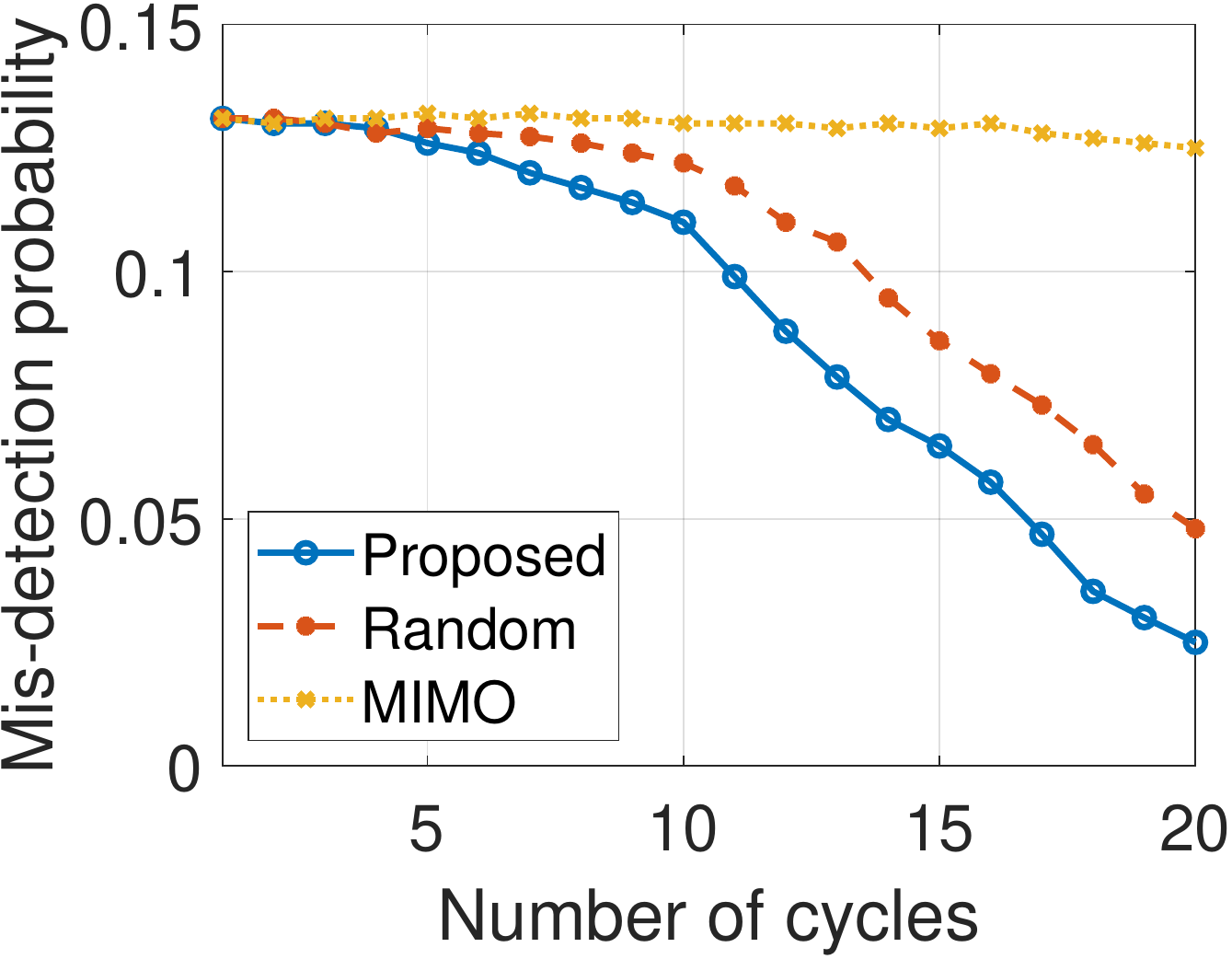}
	}
	\caption{Simulation results: (a) Detection probability vs. the number of cycles; (b) Mis-detection probability vs. the number of cycles.}
	\label{f_sim}
	%\vspace{-6mm}
\end{figure*}

In this part, the number of targets $R$ is unknown but it falls in a known range. The space of interest (SOI) is discretized into multiple angular blocks and each target is located in one block. Given $R$ and the direction of each target, the range of each target can be estimated based on the received signals. Therefore, a hypothesis $U_i$ only needs to contain the number of targets and the directions of these targets indicating the index of the angular block. The multi-target detection is performed using the multiple hypotheses testing techniques. Similar to the protocol introduced in the previous part, the MetaRadar system is also operated in a synchronized manner and the timeline is slotted into cycles. In each cycle the following three steps are performed.

\textbf{Optimization:} The aim of the optimization step is to improve the detection performance by optimizing the radar waveforms and the RIS configuration. The detection performance can be quantified by the ``distance" between the probability functions of two different hypotheses. With the RIS, the received signals $\bm{y}$ is manipulated to maximize the distance between two hypotheses, and thus any two hypotheses are more likely to be distinguished, leading to a higher detection accuracy. To be specific, the distance between hypotheses $j$ and $j'$ in the $c$-the cycle can be defined as the relative entropy~\cite{LWYQJ-2018}, i.e.,
\begin{equation}
	\begin{array}{ll}
	d^c_{j,j'}(\mathcal{P}^c) = &KL(p^c(\bm{y}|U_j,\mathcal{P}^c),p^c(\bm{y}|U_{j'},\mathcal{P}^c)) \\
	&+ KL(p^c(\bm{y}|U_j,\mathcal{P}^c),p^c(\bm{y}|U_{j'},\mathcal{P}^c)),
	\end{array}
\end{equation}
where $KL(\cdot)$ is the Kullback-Leibler divergence, $\mathcal{P}^c$ is the set of optimization variables, i.e., radar waveforms and RIS configuration, in the $c$-th cycle, and $P^c(\bm{y}^c|U_j, \mathcal{P}^c)$ is the probability to receive signal $\bm{y}^c$ given hypothesis $U_j$ in the $c$-th cycle, which can be expressed as
\begin{equation}\label{probability}
	p^{c}(U_j,\mathcal{P}^c) = A \prod \limits_{i = 1}^c \exp\left(\frac{\|\bm{y}^i - \bar{\bm{y}}^i(U_j,\{\hat{\tau}^{j}_r\}_{r = 1}^{R},\hat{\bm{\gamma}}^j,\mathcal{P}^c)\|^2}{\sigma^2}\right). 
\end{equation}
Here, $A$ is a scaling factor, $\sigma^2$ are the noise power, and $\bar{\bm{y}}^i(U_j,\{\hat{\tau}^{j}_r\}_{r=1}^R,\hat{\bm{\gamma}}^j,\mathcal{P}^c)$ denotes the  expectation of signals received by the antenna array under hypothesis $U_j$ in the $i$-th cycle, delays from each target $\hat{\tau}^{j}_r$, and their responses $\hat{\bm{\gamma}}^j$, where $\hat{\tau}^{j}_r$ and $\hat{\bm{\gamma}}^j$ can be estimated jointly using the maximum likelihood estimation method.

To solve this problem efficiently, we first decouple the problem into two subproblems: radar waveform optimization subproblem and RIS configuration optimization subproblem. For the radar waveform optimization subproblem, this problem can be transformed into a quadratically constrained quadratic program~(QCQP), which can be solved by the semidefinite relaxation (SDR) technique. For the RIS configuration optimization subproblem, the problem can also be transformed into a QCQP after relaxing the discrete phase shifts into continuous ones. After solving the QCQP, all the phase shifts will be recovered to the nearest available phase shift. More details can be found in \cite{HHBKZL-2021}.

\textbf{Transmission and Reception:} The optimized radar waveforms are transmitted. The RIS phase shifts are set as optimized in the previous step and could be different for transmission and reception modes. Then, the antenna array listens for the echo signals from the targets. Since the distances between the targets and the radar can be different, the echo signals from different targets may have different delays. The Rx will record these signals including the delay information for further processing.

\textbf{Detection:} Within this step, the probability of each hypothesis will be updated. The prior probability distribution of these hypotheses is initialized to be uniform, where $p^{1}(U_j)$ is the initial probability for hypothesis $U_j$. Based on Bayes' theorem, the probability update in the next cycle can be written as
\begin{equation}
	p^{c+1}(U_j) = \frac{p^1(U_j)P^c(\bm{y}^c|U_j)}{\sum\limits_{j'}p^1(U_{j'})P^c(\bm{y}^c|U_{j'})},
\end{equation}
where $P^c(\bm{y}^c|U_j)$ is the probability given in (\ref{probability}). At the end of the iteration, the hypothesis with the highest probability will be selected as the final result.

In Fig.~\ref{f_sim}, we show the detection probability and the mis-detection probability versus the number of cycles, respectively. In comparison, we also present the results obtained by the traditional MIMO radar and the RIS radar with a random RIS configuration. It can be observed that the detection probability obtained by the proposed scheme is higher, while the mis-detection probability is smaller than those obtained by the other two schemes, which verify the effectiveness of the proposed scheme. Moreover, we can also observe that the random and proposed schemes outperform the MIMO scheme. Specifically, the growth rate of the detection probability obtained by the MIMO scheme is much lower than that of the others, and the detection probability obtained with the RIS can approach $1$ after sufficient number of cycles. Similarly, the  mis-detection probabilities obtained by the random and proposed schemes drop significantly faster than that of the MIMO one. This has verified that incorporating the RIS can promote that performance of radar systems even the phase shifts of the RIS are not optimized by providing extra paths.

\section{RIS-aided Localization}
\label{Localization}
In this section, we will introduce the RIS-aided RF localization applications. Similar to the sensing applications, it is the main challenge to optimize the configurations of the RIS, especially when the size of the RIS is large. Based on the availability of the prior information of the environment, the techniques can be broadly categorized into two types: MetaLocalization and MetaSLAM, which will be elaborated in Section \ref{secsub:localization} and Section \ref{secsub:SLAM}, respectively.

\subsection{MetaLocalization: Indoor Localization and Tracking}
\label{secsub:localization}
In this part, we will introduce how to localize/track users if the prior information of the environment can be known. For simplicity, we use the RSS as the measurement metric. In general, an RSS-based fingerprinting system consists of two phases: \emph{offline} and \emph{online} phases. In the offline phase, the system will collect the RSS value for each sampling location and generate a radio map. Then, in the online phase, the system will estimates the user's localization by comparing its measured RSS value with the radio map \cite{ZCY-2012}. In an uncontrollable radio environment, as the radio map is passively measured, the RSS values for two neighboring locations might be similar to each other, leading to a performance degradation. To address this issue, the RIS is used in an RF localization system to actively alter the radio maps and reduce the similarity of the RSS values corresponding to two adjacent locations. Such a system is referred to as MetaLocalization~\cite{HHBKZL-2021-TWC}.

\begin{figure}[t]
	\centering
	\includegraphics[width=2.8in]{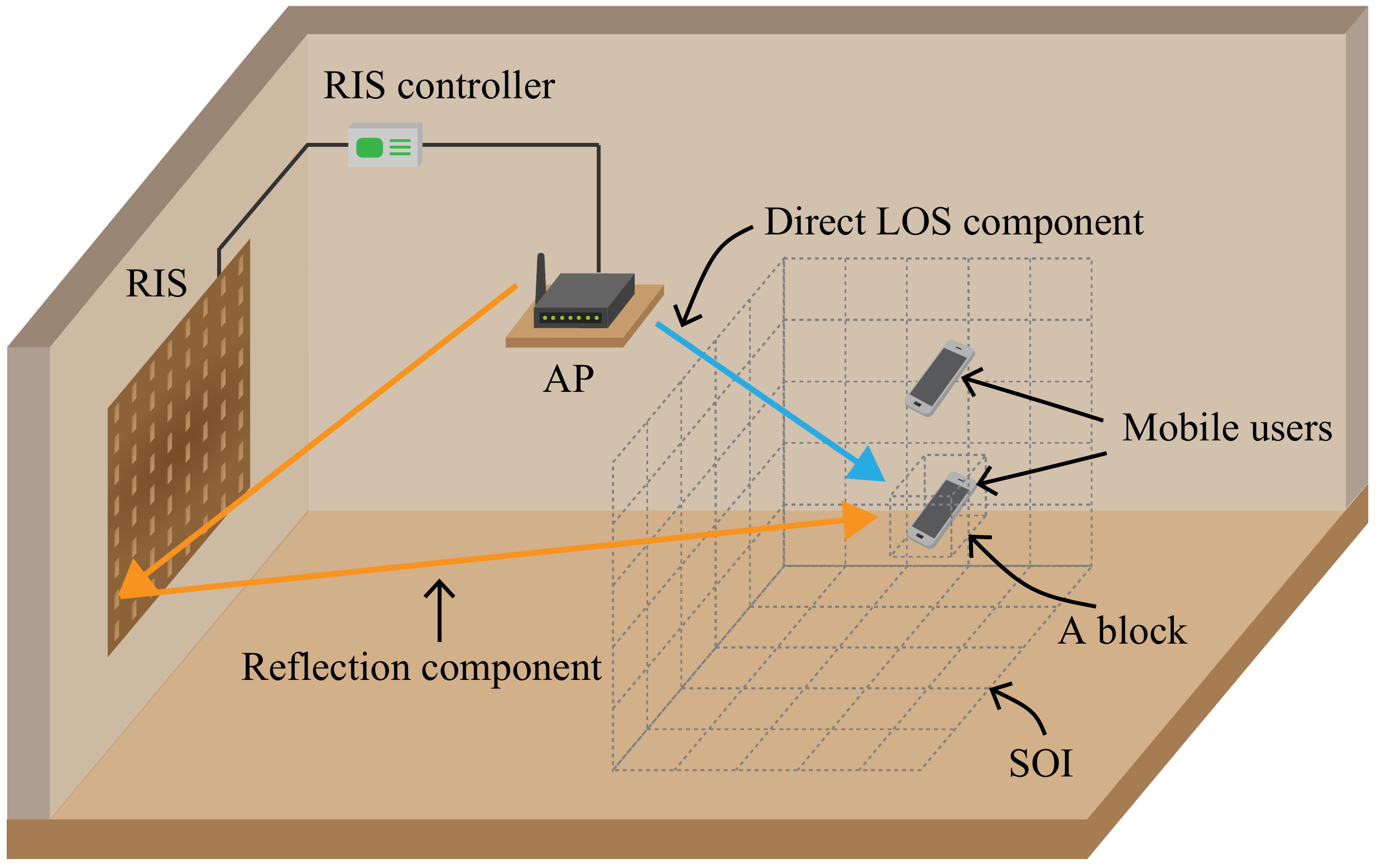}
	\caption{Illustration for MetaLocalization Systems.}
	\label{MetaRadar:scenario}
\end{figure}

The MetaLocalization system is illustrated in Fig.~\ref{MetaRadar:scenario}. The system is composed of an AP, an RIS, and multiple users requiring indoor location services. The AP connects to the RIS controller to facilitate the synchronization. During the localization process, the AP sends a single-tone signal, and the RIS reflects the signal to users. Then, each user measures the RSS for localization. All the mobile users are assumed to move slowly or stay static in the SOI, which is divided into several cubic blocks with the same size. The location of each user can be represented by the index of the block. The localization process has several cycles, each consisting of two phases: the radio map generation phase and the localization estimation phase, as elaborated below.

\subsubsection{Radio Map Generation Phase}
To improve the localization accuracy, the MetaLocalization system needs to adjust configurations of the RIS to reconfigure the radio environment and provide a favorable radio map. To be specific, for a certain configuration, the RSS for each block in the SOI can be calculated according to (\ref{RSS}), and we can obtain the radio map by repeating this step for all the configurations. However, as the number of available configurations can be very large with a large size of the RIS, it is costly to measure the RSS values in the SOI for all the configurations. 

To address this issue, we need to select a configuration which can lead to the minimum localization loss in each cycle. More specifically, the localization loss can be defined as the sum of expected localization errors for all the users \cite{HHBKZL-2021-L}, which can be expressed as
\begin{equation}
	l(\bm{c}, \mathcal{L}) = \sum_{i \in \mathcal{S}}\sum_{\substack{q,q' \in \mathcal{Q}\\ q\ne q'}} p_{i, q} \gamma_{q, q'}  \int p (s_i | \bm{c}, q) \mathcal{L}(q' | \bm{c}, s_i) ds_i,
	\label{def_loss}
\end{equation}
where $\mathcal{S}$ is the set of users, $\mathcal{Q}$ is the set of blocks in the SOI, and $\bm{c}$ is the configuration of the RIS. Here, $p_{i,q}$ is the prior distribution of the users over the SOI, which can be obtained by the results in the previous iteration. $\gamma_{q, q'}$ is the weight for the mislocalization, which is defined as the Euclidean distance between the ground-truth block $q$ and the estimated block $q'$. This will force the estimated location to get closer to the ground-truth one as much as possible. $p(s_i|\bm{c},q)$ denotes the probability of the received RSS for user $i$ being $s_i$ with configuration $\bm{c}$ when user $i$ is located at the $q$-th block, and $\mathcal{L}(q' |\bm{c}, s_i)$ is an estimation function to indicate whether user $i$ is located at the $q'$-th block with the received RSS $s_i$, which is assumed to be known in this phase and will be discussed in the following phase. Therefore, the integration is the probability that the system estimates user $i$ to be located at the $q'$-th block while it is actually located at the $q$-th block.

The loss function is non-convex with respect to the RIS configuration $\bm{c}$. Moreover, the configuration of the RIS is typically discrete in practice. To address this problem efficiently, we can first find some initial solutions and use the global descent methods to update the solution leading to a lower localization cost. The details of the algorithm can be found in \cite{HHBKZL-2021-TWC}. 

\subsubsection{Location Estimation Phase} Given configuration $\bm{c}$ and the RSS for each user $s_i$, the optimal estimation function which yields to the minimum localization loss can be expressed as \cite{HHBKZL-2021-TWC}
\begin{equation}\label{estimation}
	\mathcal{L}^*(q' | \bm{c}, s_i) =
	\begin{cases}
		1, & s_i \in \mathcal{R}_{i, q'},\\
		0, & s_i \notin \mathcal{R}_{i, q'},
	\end{cases}
\end{equation}
where the decision region $\mathcal{R}_{i, q'}$ is defined as
\begin{align}
		\mathcal{R}_{i, q'} = \bigg\{& s_i : \sum_{q\in\mathcal{Q}} p_{i, q} (\gamma_{q,q'} - \gamma_{q,q''}) p (s_i | \bm{c}, q) \le 0, \notag\\
		&\forall q'' \in \mathcal{Q}/\{q'\}\bigg\}.\label{pro_opt_decision_c1}
\end{align}
The received RSS falling within the decision region implies that the localization loss will be less if we estimate that user $i$ is located at the $n'$-th block instead of the $n''$-th block. Therefore, in this phase, each user can estimate its location using (\ref{estimation}) according to the configuration $\bm{c}$ and the received RSS.

\begin{figure}[!t]
	\centering
	%\label{a1} %% label for first subfigure
	\includegraphics[width=2.8in]{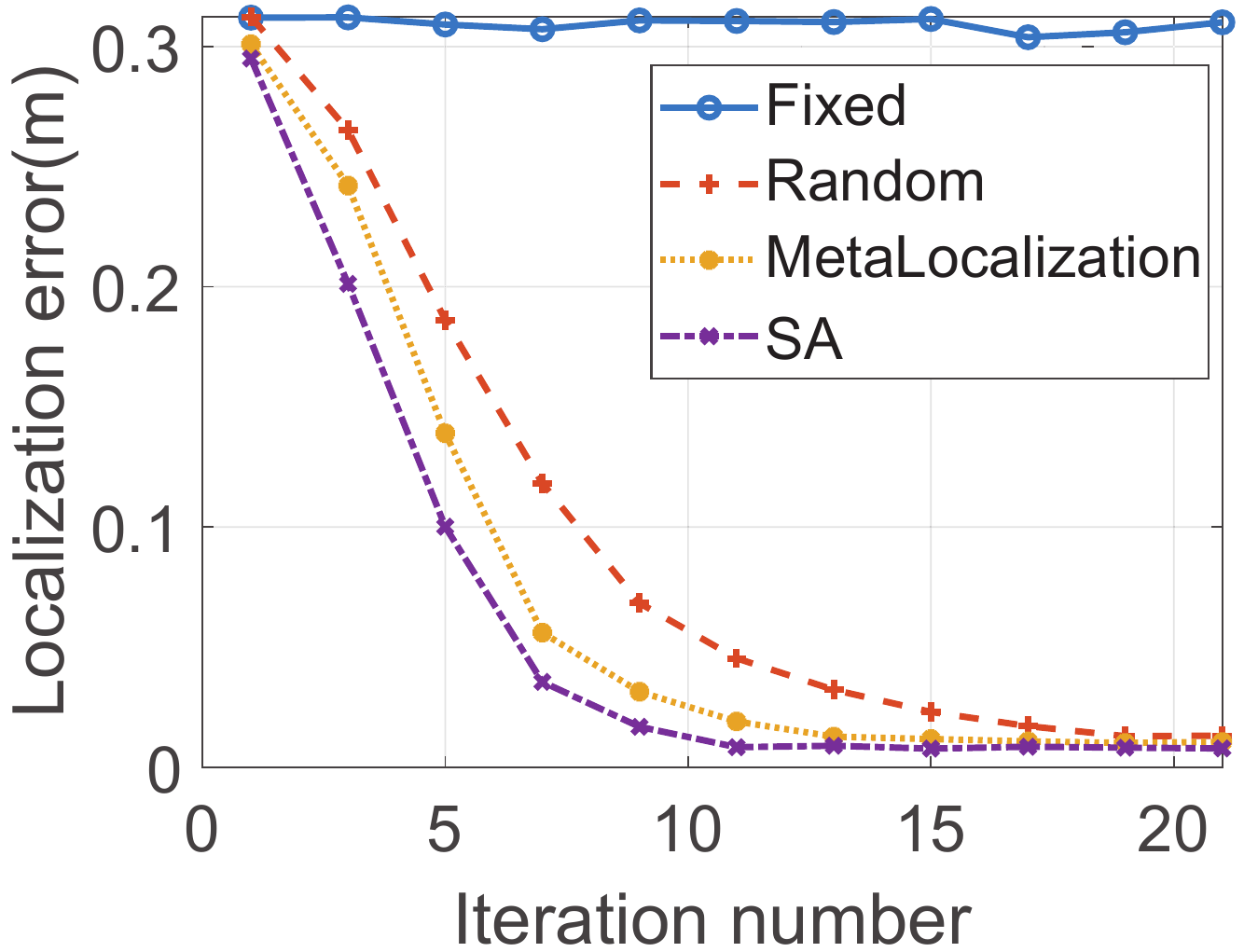}
	\caption{Localization performance for a single user.}
	\label{f_single}
\end{figure}

Fig.~\ref{f_single} illustrates the localization performance for a single user. To evaluate the performance of the MetaLocalization scheme, we also give the performance obtained by another three schemes: the fixed configuration scheme, the random configuration scheme, and the simulated annealing (SA) scheme. In the fixed configuration scheme\footnote{It is worth pointing out that the fixed configuration scheme can represent the case without the RIS. When the phase shifts of the RIS are fixed, the RIS can act as a normal wall which scatters the signals.}, the states of all the RIS elements are fixed.  In the random configuration scheme, random configurations are generated in different iterations. In the SA scheme, the simulated annealing method is utilized to optimize the RIS configurations, which can be regarded as a lower bound. We can observe that the localization error obtained by the fixed algorithm fluctuates between $0.30$ m and $0.32$ m, while that obtained by other three schemes decrease when the number of iterations increases. Moreover, we can observe that the localization error of the MetaLocalization scheme is close to that obtained by the SA scheme, can achieve cm-level localization accuracy. This indicates that the capability of the RIS to customize the propagation environment can improve the localization performance.  

\subsection{MetaSLAM: Simultaneous Localization and Mapping}
\label{secsub:SLAM}
In the previous part, we have introduced how to determine the users' locations if we have a prior information of the environment. However, in some applications, for example, a mobile robot is placed at an unknown environment to execute a certain task, how can this robot incrementally build a consistent map of this environment and locate itself within this map? This arises to the simultaneous localization and mapping~(SLAM) technique \cite{HT-2006}. In the radio-based SLAM technique, the multipath propagation caused by scatters in the environment is exploited, and thus the locations of objects in the environment (scatter points) are simultaneously determined with the agent's location \cite{CTWSAU-2016}. Even if the multipath channel is used as a constructive source of information in the localization problem, the related electromagnetic (EM) interactions with the environment still remain uncontrolled and, as such, largely suboptimal from the localization perspective \cite{HJBAM-2020}. This motivates us to leverage the RIS to improve the accuracy of the SLAM technique.

\subsubsection{Indoor SLAM Scenario}
\begin{figure}[t]
	\centering
	\includegraphics[width=2.8in]{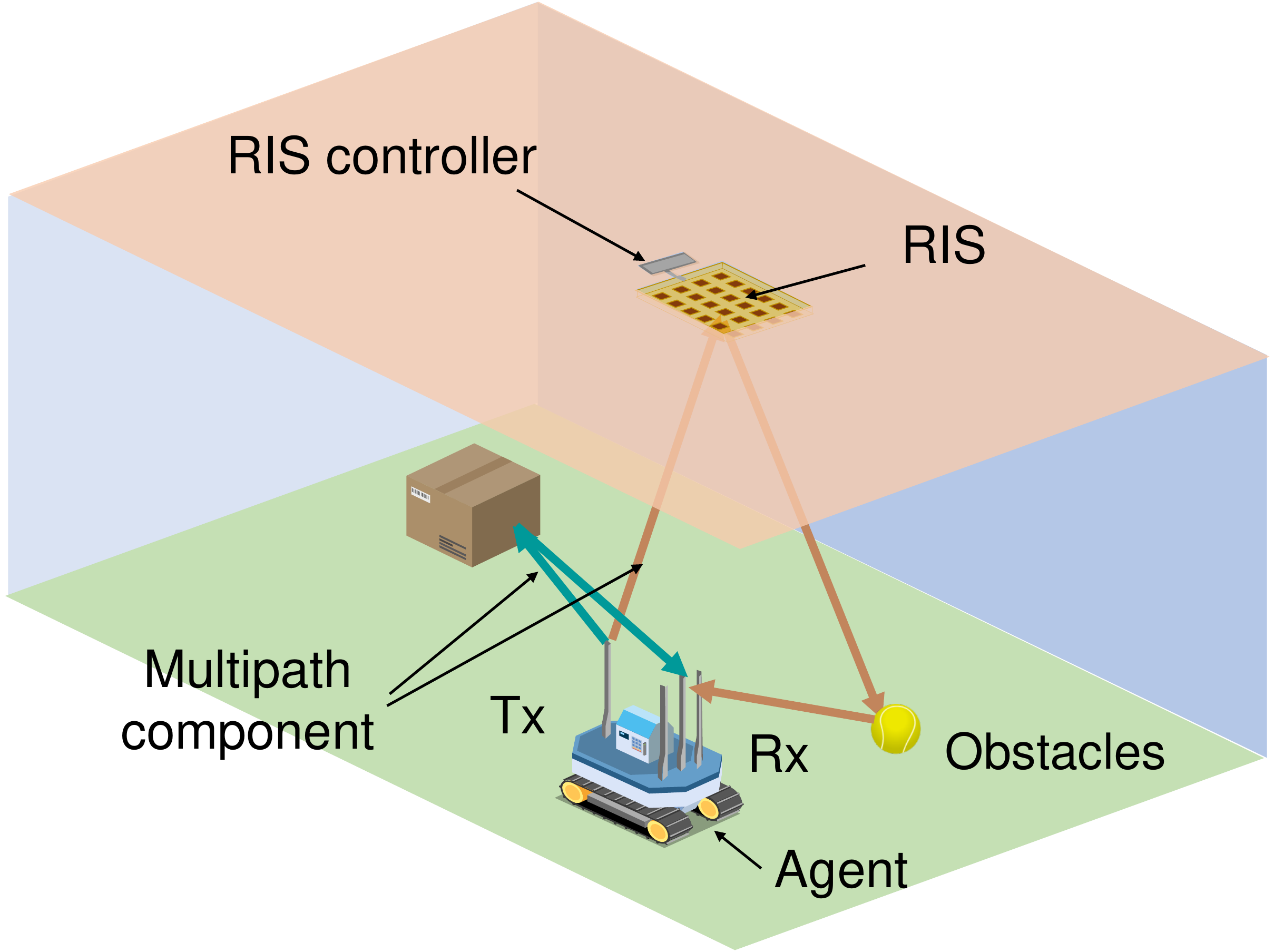}
	\caption{An illustration for an indoor RIS-assisted SLAM scenario.}
	\label{SLAM:scenario}
\end{figure}

A general indoor RIS-assisted SLAM scenario is illustrated in Fig. \ref{SLAM:scenario}. An RIS is deployed on the ceiling of a room, connecting to a controller that can change the phase shifts applied by the RIS to customize the propagation environment. A mobile agent is equipped with a single-antenna Tx and a multiple-antenna Rx. While moving in the room, the agent transmits signals and analyzes the received signals in order to locate itself and map the surrounding environment simultaneously. In particular, the agent first communicates with the RIS controller to adjust the phase shifts of the RIS. Next, the agent simultaneously emits signals to the environment and records the received signals, which contain multipath components produced by the scattering and reflection of environmental obstacles and the RIS. The location of the mobile agent and environmental information can be extracted from these multipath components~\cite{ZHBHKL-2021}. 

\subsubsection{Multipath Component Modeling}

\begin{figure}[t]
	\centering
	\includegraphics[width=2.8in]{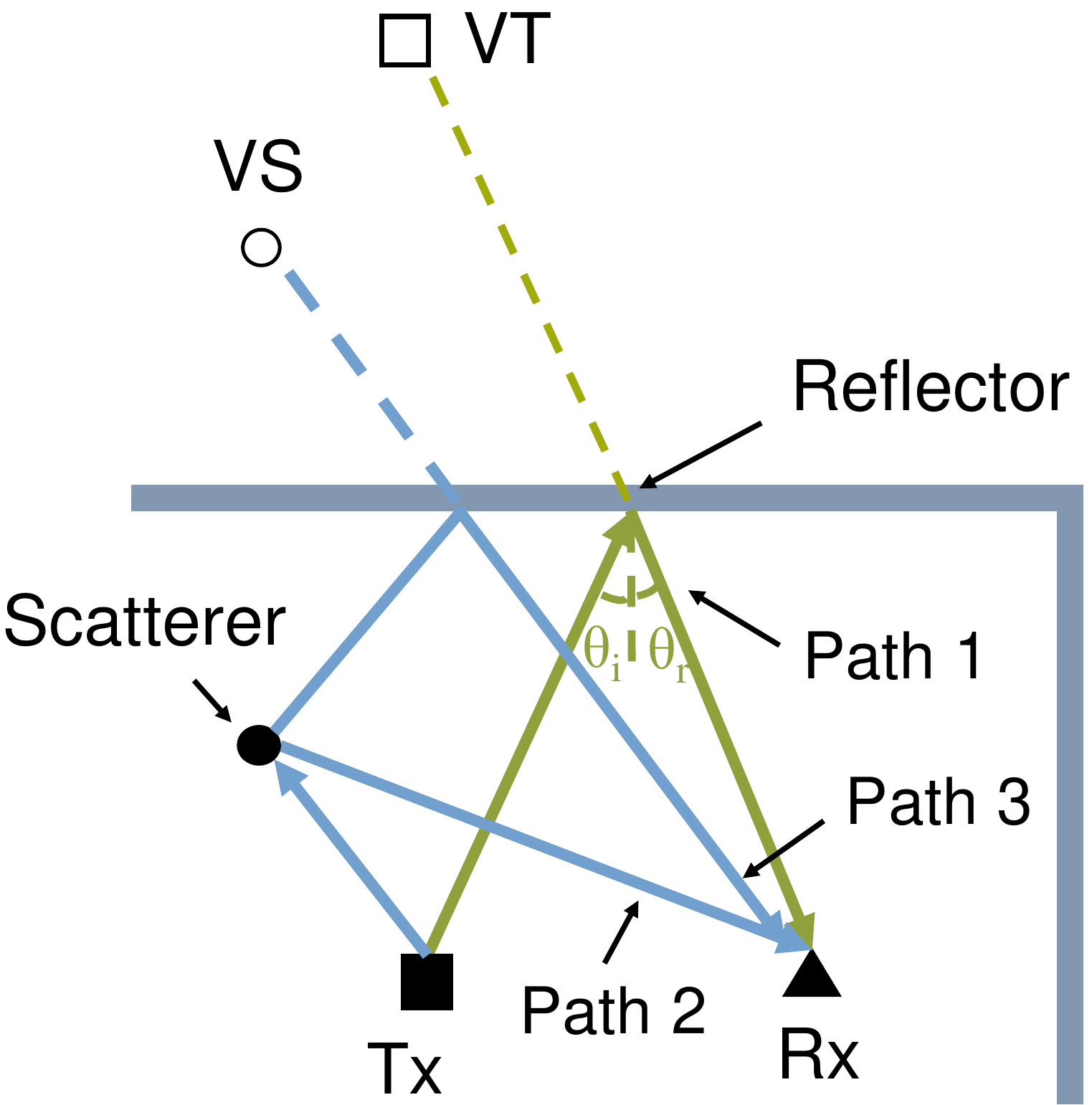}
	\caption{An example for the multipath channel.}
	\label{SLAM:multipath}
\end{figure}

Assume the transmitted OFDM symbol $x^n(t)$ over the $n$-th subcarrier reaches the $v$-th Rx antenna via $L$ paths. Therefore, the baseband OFDM symbol after removing the CP received by the $v$-th Rx antenna at time $t$ can be written as
\begin{equation}
\label{eq:all}
y^n_v(t)=\sum_{l=1}^{L} \int_{-\infty}^{\infty} \hat{h}^n_{l,v}(\tau)x(t - \tau) d\tau + \omega_v^n(t),
\end{equation}
where $\hat{h}^n_{l,v}(t)$ denotes the channel impulse response (CIR) of the $l$-th multipath channel to the $v$-th Rx antenna over the $n$-th subcarrier, and $\omega_v^n(t)$ denotes the Gaussian white noise.

In the following, we model the impulse response of channels via the obstacles and the RIS. Two types of obstacles are considered: reflectors and scatterers. The reflectors are smooth surfaces like walls which receives a section of the incident wavefront and redirects it following the reflection law, while the scatterers only receives a point on the incident wavefront and then diffusely scatter it in all directions around themselves. An example for the multipath channel is illustrated in Fig. \ref{SLAM:multipath}.

\textbf{Path through a reflector:} As illustrated in Fig.~\ref{SLAM:multipath}, path 1 is the signal path from the Tx to the Rx via a reflector on the ceiling. The CIR of path~1 in general can be expressed as
\begin{equation}
	\label{a1}
	\hat{h}^n_{1,v}(t)=h^n_{1,v}\delta(t-d_1/c),
\end{equation}
where $h^n_{1,v}$ is the channel gain as introduced in (\ref{LoS_Channel}), and $d_1$ is the propagation length of this path, and $\delta(\cdot)$ is the delta function. As illustrated in Fig.~\ref{SLAM:multipath}, this CIR for this path is equivalent to that from a virtual Tx (VT), which is the mirror image of the Tx. This presentation will be used in the following algorithm design.

\textbf{Path through a scatterer:} As illustrated in Fig.~\ref{SLAM:multipath}, path~2 is the signal path from the Tx to the Rx via a scatterer. Its CIR can be expressed as that in (\ref{a1}), where the channel gain will be changed as defined in (\ref{RIS_Channel}). It is worthwhile to point out that each RIS element can also be regarded as an scatterer with a controllable delay. If the scatter is not an RIS element, $\gamma_m$ should be set as 1 when calculating the channel gain according to (\ref{RIS_Channel}).

\textbf{Path through a scatterer and a reflector:} Path~3 is the propagation path from the Tx to the Rx via a scatterer and a reflector. Similarly, path 3 can be treated as being transmitted by a virtual scatterer (VS) with an additional delay introduced by the transmission between the Tx and the scatterer. The VS is also the mirror image of the scatterer, which will be used in the following algorithm design as well.

\subsubsection{Position and Mapping Procedure}
In the positioning and mapping procedure, we similarly divide the timeline into cycles. In each cycle, the following three steps are conducted sequentially.

\textbf{Step 1 Phase Shift Optimization:} In this step, the agent needs to select the phase shifts of the RIS at current cycle. The objective of this optimization problem is to minimize the localization error in the current cycle, i.e., the distance between the ground-truth and estimated locations. However, since the ground-truth location of the agent is unknown, it is difficult to optimize the phase shifts of the RIS with the positioning error as the objective. Alternatively, we use the Cramer-Rao lower bound~(CRLB) to approximate the positioning error, which is widely used in the performance measurement of a SLAM system. In particular, the CRLB is reciprocal to the Fisher information matrix of the estimated location \cite{KP-2012}. As the CRLB is non-convex with respect to the phase shifts of the RIS, a genetic algorithm can be used to solve this problem~\cite{ZHBHKL-2021}.

\textbf{Step 2 Communication and Measurement:} After Step 1, the agent transmits a  signal over a control channel carrying the phase shifts of the RIS to the RIS controller. The RIS controller will adjust the phase shifts accordingly. Once the phase shifts of the RIS are updated, the agent will transmit another signals for SLAM and the Rx records the received signals at the same time.

\textbf{Step 3 Localization and Mapping:} The location of the agent and the map is updated based on the signals measured in Step 2. The localization and mapping procedure consists of two phases, i.e., path grouping phase and positioning and mapping phase. 

\begin{itemize}

\item \emph{Path Grouping Phase:} In this phase, we first need to recognize whether the received path is from a scatterer or a VS, since scatterers and VSs are static. The Tx or VTs are moving during the SLAM process, and thus cannot be used to locate the agent. The recognition can be achieved by a neural network \cite{SWD-2005}. Next, these paths are divided into several groups based on their AoAs, each corresponding to a scatterer (including the RIS), which is also referred to as \emph{landmark}. The VS can be mapped with a scatterer according to the geometry. Finally, we need to decide which landmark is the RIS, which is necessary for the phase shift optimization. Let $p^c_{i, RIS}$ be the probability that the $i$-th landmark is the RIS in the $c$-th cycle, and we regard the probability obtained by the previous cycle as the prior distribution for the current cycle. Based on the Bayes theorem, we have
\begin{equation}
	p^c_{i,RIS}=\frac{p_{i,RIS}^{c-1}p(\bm{a}^c|i\text{-th landmark is~the RIS})}{\sum_{i=1}^Np_{i,RIS}^{c-1}p(\bm{a}^c|i\text{-th landmark is~the RIS})},
\end{equation}
where $\bm{a}^c$ denotes the amplitudes of paths extracted from the received signal $y^c$, $p(\bm{a}^c|i \text{-th landmark is~the RIS})$ denotes the probability to receive $\bm{a}^c$ if the $i$-th landmark is RIS. The \emph{landmark} with the largest $p_{i,RIS}^c$ will be regarded as the RIS in the $c$-th cycle.

\item \emph{Positioning and Mapping Phase:} The positioning and mapping algorithm is based on the particle filter method~\cite{HT-2006}. The basic idea is to use a set of weighted particles to represent the probability distribution of the locations of the agent and \emph{landmarks}, and the weights of these particles are updated based on the received signals. Since the estimation errors of ToFs and AoAs vary among different paths, we will optimize the weights of these paths in the particle filter to reduce the positioning errors. It is worth pointing out that the computational complexity increases exponentially with the number of landmarks held in the map. We can use partitioned updates and relative submaps to address this issue \cite{TH-2006}.

\end{itemize}

\begin{figure}[t]
	\centering
	\includegraphics[width=3.0in]{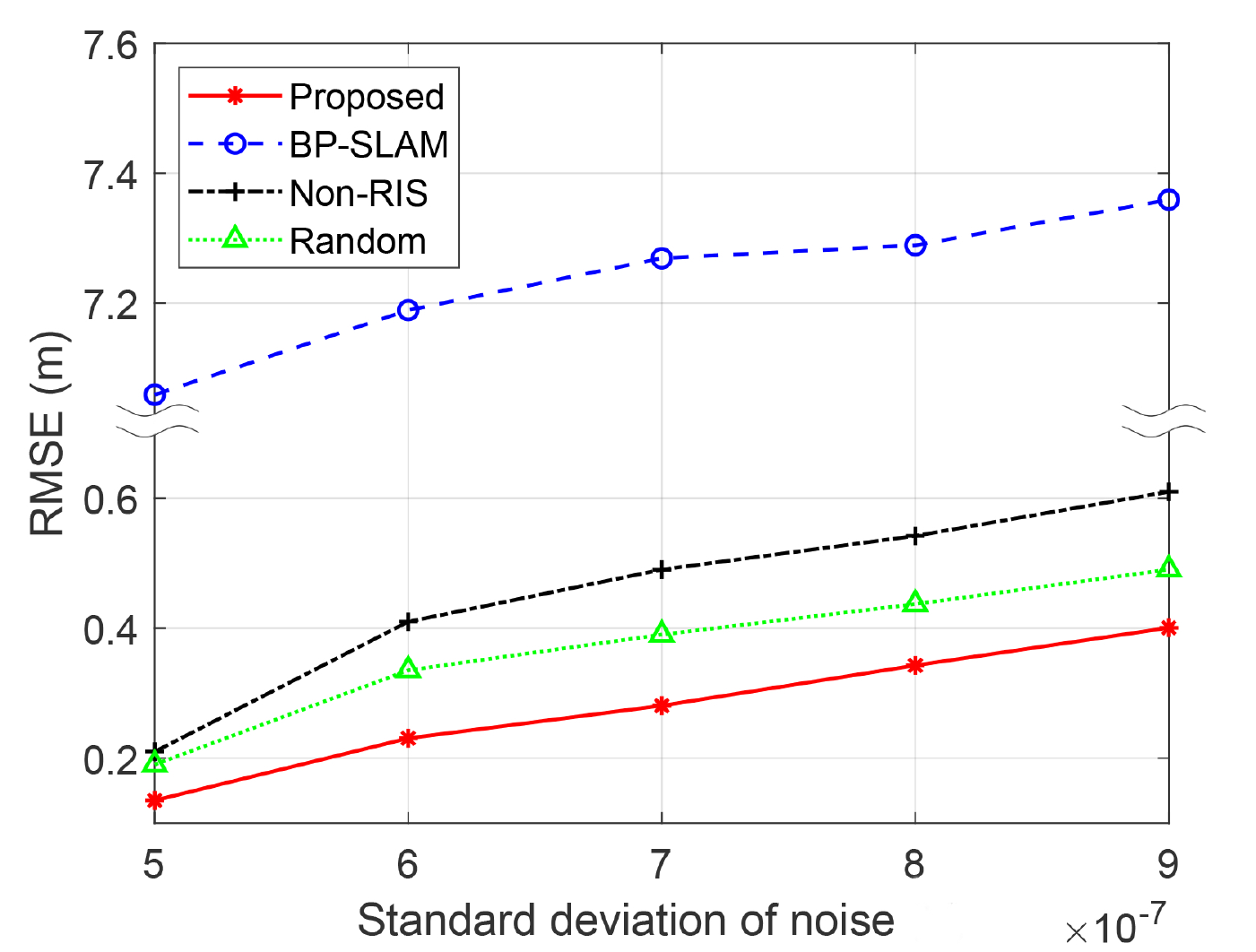}
	\caption{Agent position RMSE vs. standard deviation of noise.}
	\label{SLAM:simulation}
\end{figure}

Fig.~\ref{SLAM:simulation} shows the root mean square error (RMSE) of the estimated agent position vs. the standard deviation of noise. In comparison, the simulation results of the following three schemes are also provided: 1) \emph{Random scheme}: The phase shifts for the RIS are set randomly in each cycle; 2) \emph{Non-RIS SLAM}: There is no RIS in the room, and the agent performs SLAM using the proposed localization and mapping algorithm; 3) \emph{BP-SLAM}: This algorithm is proposed in \cite{EFFKFM-2019}, which only utilizes the TX and VTs for localization and mapping. From this figure, we can observe that the RMSE of the agent position obtained by the proposed scheme is at least 31\% lower than other benchmark schemes, which shows the superiority of the proposed SLAM scheme. Moreover, compared to those without the RIS, the RMSEs obtained by the RIS-aided SLAM systems are lower, implying that the effectiveness of the RIS to improve the performance of SLAM systems.

\section{System Implementation and Experimental Evaluation}
\label{Implementation}

In Section \ref{subsec:component}, we will introduce how to build a RIS-aided wireless sensing and localization hardware system, and in Section \ref{subsec:performance}, we show some results obtained in the hardware systems.

\subsection{System Components}
\label{subsec:component}

\begin{figure}[!t]
	\centering
	\includegraphics[width=3.0in]{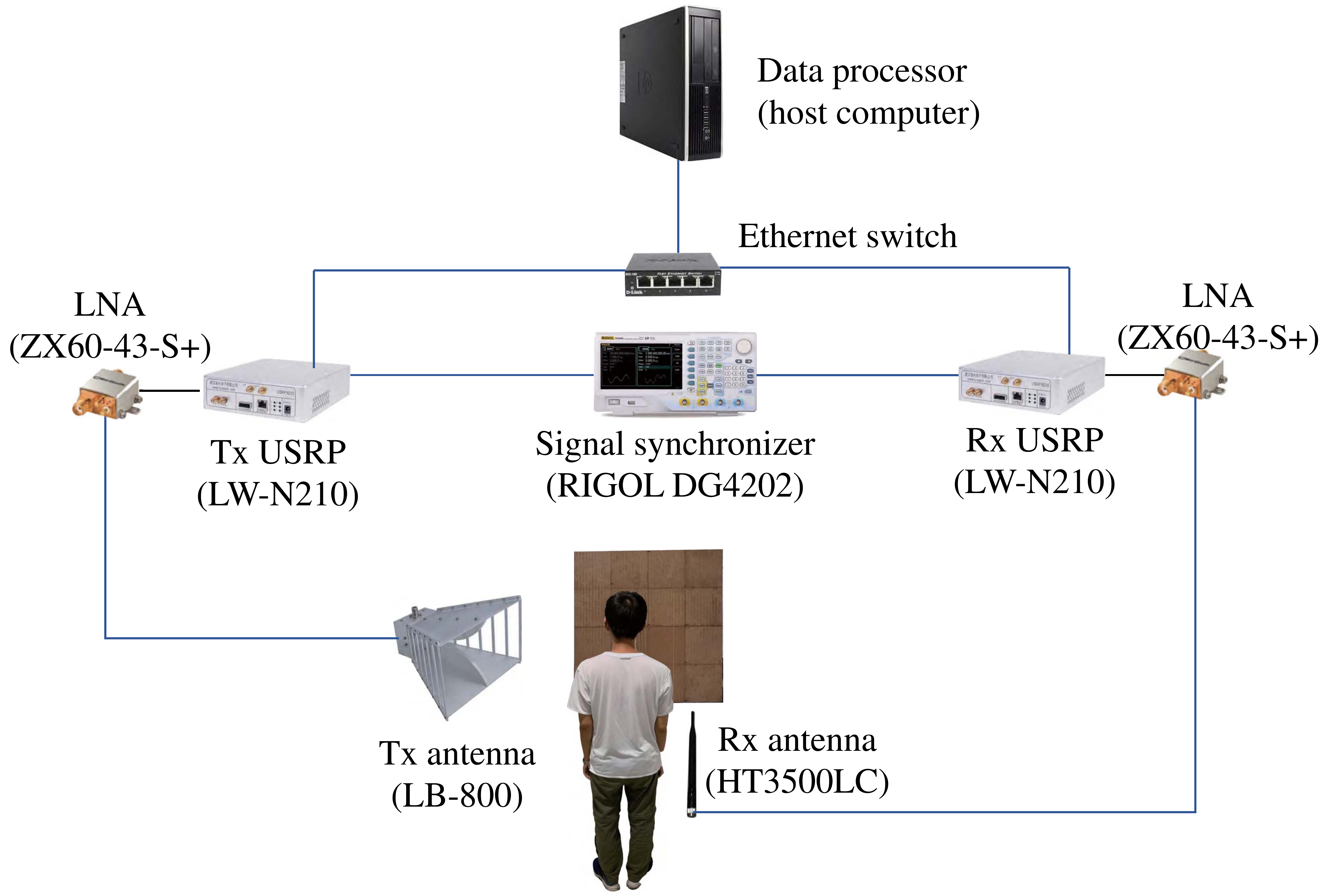}
	%	\vspace{-2em}
	\caption{An example prototype for RIS-aided wireless sensing and localization systems.}
	\label{fig: transceiver}
\end{figure}

An example prototype of an RIS-aided wireless sensing and localization system is illustrated in Fig. \ref{fig: transceiver}. In general, the system composes of two parts: the RIS and transceiver modules, which are elaborated as follows.

\subsubsection{RIS Implementation}

\begin{figure}[!t]
	\centering
	\includegraphics[width=3.0in]{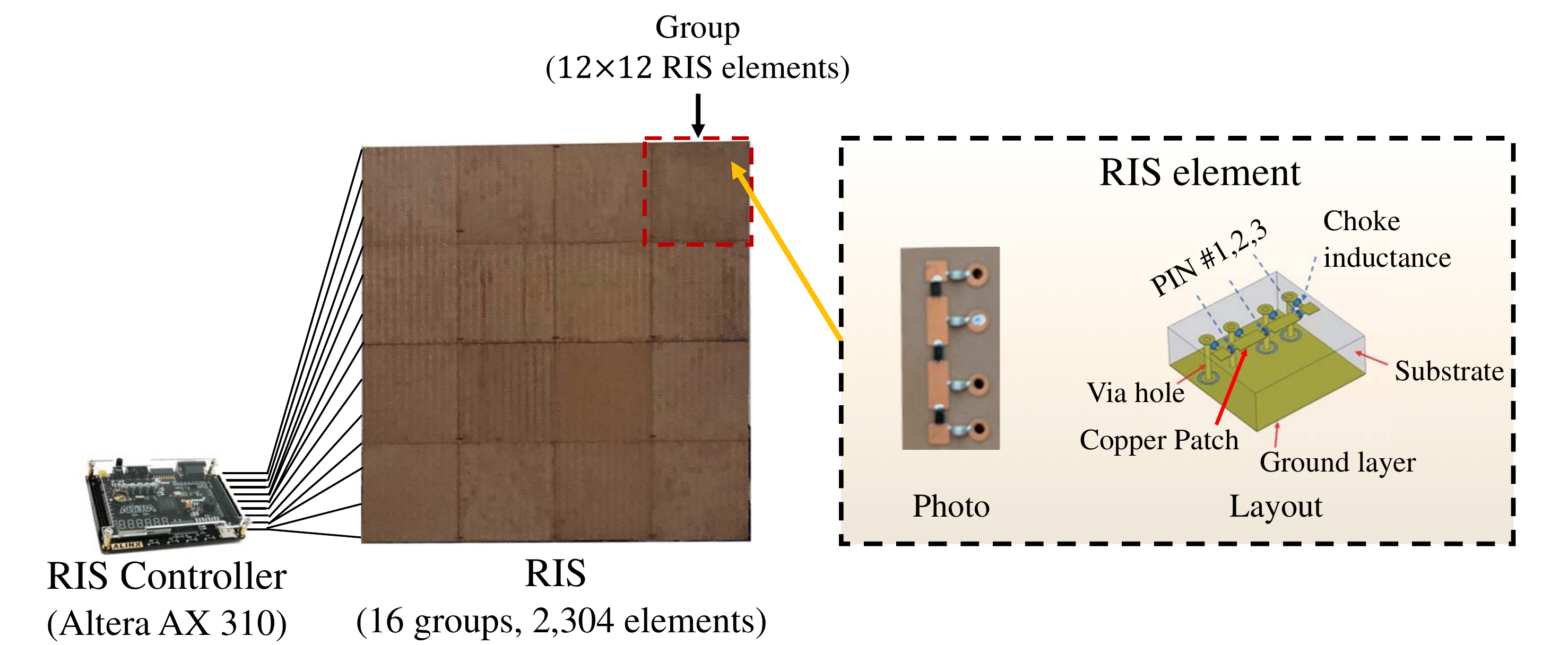}
	\caption{RIS controller and RIS.}
	\label{fig: RIS}
\end{figure}

We use the electrically modulated RIS proposed in~\cite{LHCYYACT-2019}, which is shown in Fig.~\ref{fig: RIS}. The RIS is a two-dimensional array with the size of $69\times 69\times 0.52$ cm$^3$, where each row/column of the array contains 48 RIS elements. Each RIS element has a size of $1.5\times 1.5 \times 0.52$ cm$^3$ and is composed of 4 rectangle copper patches printed on a dielectric substrate~(Rogers 3010) with a dielectric constant of 10.2. Two adjacent copper patches are connected by a PIN diode~(BAR 65-02L), and each PIN diode has two states, i.e., ON and OFF, which are controlled by applied bias voltages on the via holes. Besides, to isolate the DC feeding port and microwave signal, four choke inductors of $30$nH are used in each RIS element. 

As shown in Fig. \ref{fig: RIS}, 3 PIN diodes are included in an RIS element, an RIS element could have at most 8 states. However, 4 states are used for the ease of control. Table~\ref{table: response} provides the amplitude and phase responses of an RIS element for these four states with incident signals of $3.198$~GHz \cite{JHBLLYZH-2020}. Table~\ref{table: response} is obtained with Microwave Studio and the Transient Simulation Package in the CST software, by assuming normal illumination. In practice, to relieve the complexity of the control circuit, the RIS elements are divided into several groups, for example, 16 groups in this prototype, each containing $12\times 12$ RIS elements arranged squarely. The RIS elements within the same group are in the same state.

\begin{table}[!t]
	\caption{Amplitude and Phase Responses of an RIS Element}
	\centering
	\scriptsize
	\begin{tabular}{| c | c| c | c | p{1cm}<{\centering}| p{1cm}<{\centering} |}
		\Xhline{1pt}
		\textbf{State} & \textbf{PIN \# 1} &\textbf{PIN \#2} &\textbf{PIN \#3} & \textbf{Phase} & \textbf{Amplitude}\\
		\hline
		$1$ & OFF & OFF & OFF & $\pi/4$ & $0.97$ \\
		$2$ & ON & OFF & ON & $3\pi/4$ & $0.97$ \\
		$3$ & ON & OFF & ON  & $5\pi/4$ & $0.92$ \\
		$4$ & ON & ON & OFF & $7\pi/4$ & $0.88$ \\
		\Xhline{1.pt}
	\end{tabular}
	\label{table: response}
\end{table}

As shown in Fig.~\ref{fig: RIS}, the states of these RIS elements are configured by an \emph{RIS controller}, which is implemented by a field-programmable gate array~(FPGA)~(ALTERA AX301). Specifically, the expansion ports on the FPGA are used for the configuration of the RIS. Every three expansion ports control the state of one group by applying bias voltages on the PIN diodes. The algorithms introduced in previous sections are loaded to the FPGA to adjust the configurations of the RIS automatically.  

\subsubsection{Transceiver Modules}
As shown in Fig.~\ref{fig: transceiver}, the transceiver module consists of the following components:
\begin{itemize}
	\item \textbf{Transmitter}: The Tx is implemented by using a universal software radio peripheral (USRP) device (LW N210). The USRP realizes the functions of RF modulation/demodulation and baseband signal processing by using a GNU Radio software development kit. The output port of the USRP is connected to a ZX60-43-S+ low-noise amplifier (LNA), which amplifies the transmitted signal. A directional double-ridged horn antenna (LB-880) is employed, which is linearly polarized.
	
	\item \textbf{Receiver}: Similar to the TX, the Rx is a USRP, whose input port is connected to an LNA, and an omni-directional vertical antenna~(HT3500LC) is utilized. An external clock (10MHz OCXO) is used to provide a precise clock signal to the Tx and Rx.
	
	\item \textbf{Signal Synchronizer}: To obtain the relative phases and amplitudes of the received signals with respect to the transmitted signals, we employ a signal source~(RIGOL DG4202) to synchronize the Tx and Rx USRPs. The signal source provides the reference clock signal and the pulses-per-second~(PPS) signal to the USRPs, which are used for the modulation and demodulation. For the phase synchronization, the Tx and Rx USRPs are connected by a wired link with a fixed gain, which is used to compensate the instrumental error of the USRPs. 
	
	\item \textbf{Ethernet switch}: The Ethernet switch connects the USRPs and a host computer to a common Ethernet, where they exchange the transmitted and received signals.
	
	\item \textbf{Data processor}: The data processor is a host computer which controls the Tx and Rx by using Python programs. The host computer also extracts and processes the received signals .
\end{itemize}

\subsection{Performance Evaluation}
\label{subsec:performance}
In this part, we will show some experimental results obtained by the shown prototype to validate the effectiveness of the proposed RIS-assisted wireless sensing and localization schemes. 

\begin{figure}[!t]
	\centering
	\includegraphics[width=2.5in]{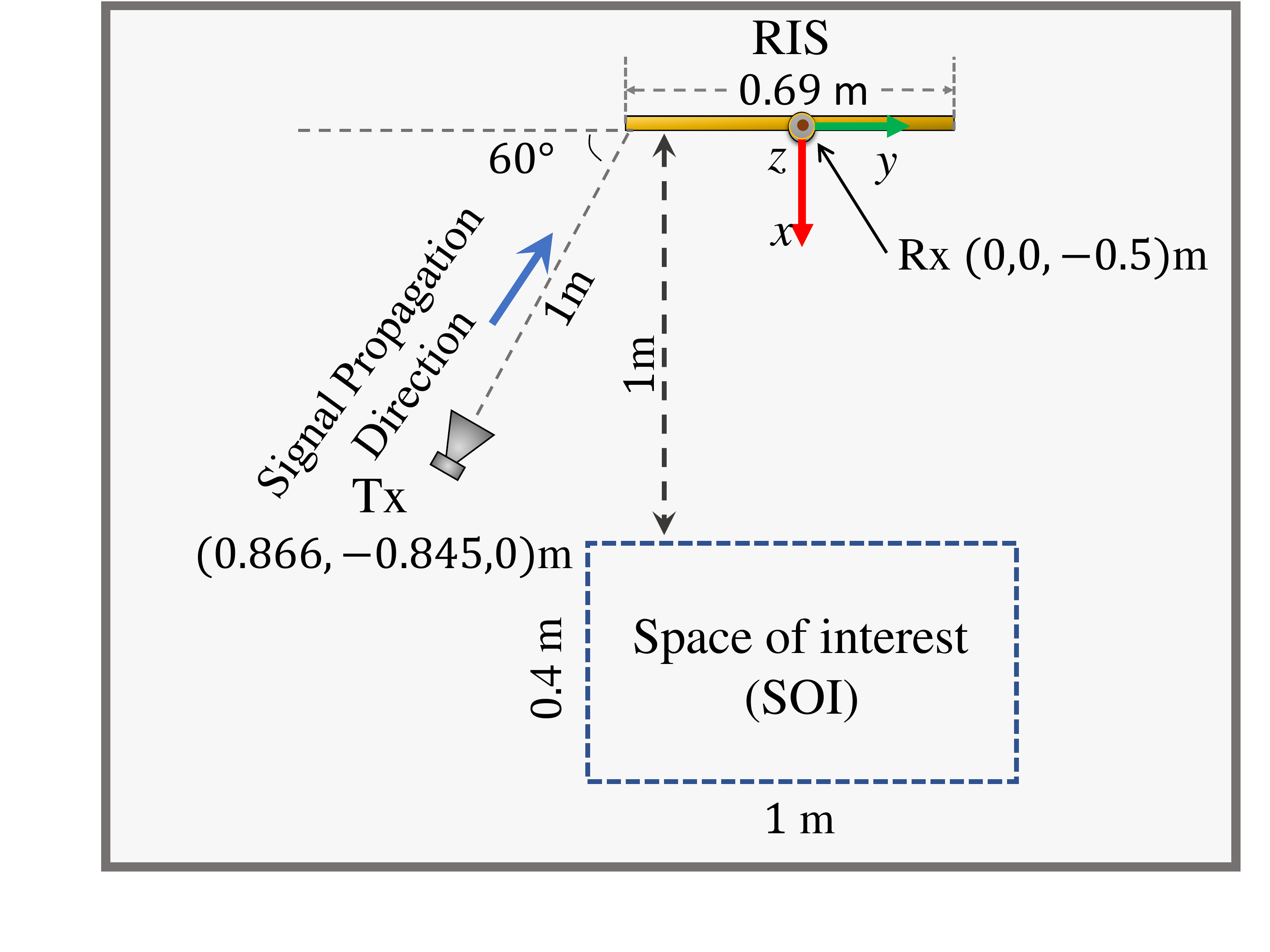}
	\caption{Experimental layout for the MetaSensing system.}
	\label{s_layout}
\end{figure}

\begin{figure*}[!t] 
	\center{\includegraphics[width=0.90\linewidth]{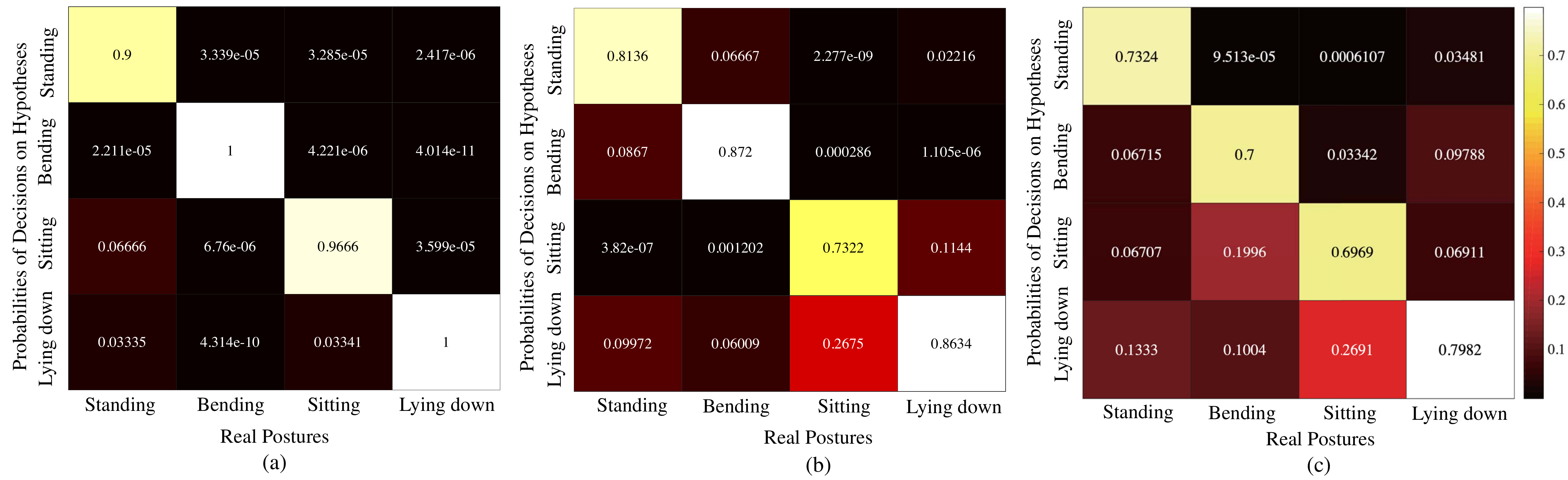}}
	%\vspace{-2em}
	\caption{Posture recognition accuracy with (a) optimized configuration; (b) random configuration; and (c) no-RIS.}
	\label{fig: posture recognition precision}
\end{figure*}

\subsubsection{RIS-aided Sensing}
For sensing applications, we show the posture recognition as an example. In this experiment, we consider 4 postures for recognition: standing, sitting, bending, and lying down. For each posture, we collect 150 labeled measurements with a random configuration and an optimized configuration of the RIS, respectively, and form the data sets. 

The layout of the experiment is shown in Fig. \ref{s_layout}. The origin of the 3D-coordinate is located at the center of the RIS, and the RIS is in the $y-z$ plane. Besides, the $z$-axis is vertical to the ground and pointing upwards, and the $x$- and $y$-axes are parallel to the ground. The Tx antenna is located 1~m away from the corner of the RIS and the Rx antenna is placed below the RIS. The human body is in the SOI, which is a cuboid region located at $1$~m from the RIS. The side lengths of the SOI are $l_x = 0.4$~m, $l_y = 1.0$~m, and $l_z = 1.6$~m. Moreover, the SOI is further divided into $M = 80$ cubics with a side length of $0.2$~m. For more detailed experimental settings, please refer to \cite{JHBLLYZH-2020}.

Fig. \ref{fig: posture recognition precision} shows the accuracy of the posture recognition with the optimized configuration, random configuration, and no-RIS, respectively. In each subfigure, the diagonal elements are the recognition accuracy for each posture. It can be observed that the system with optimized configuration can achieve $14.6\%$ higher recognition accuracy compared to that with random configuration. This verifies the necessity of the phase shift optimization. Moreover, comparing to the no-RIS case, we can observe that the RIS can significantly increases the posture recognition accuracy even with a random recognition. This justifies the effectiveness of the RIS-aided RF sensing.

\begin{figure}[!t]
	\centering
	\includegraphics[width=2.8in]{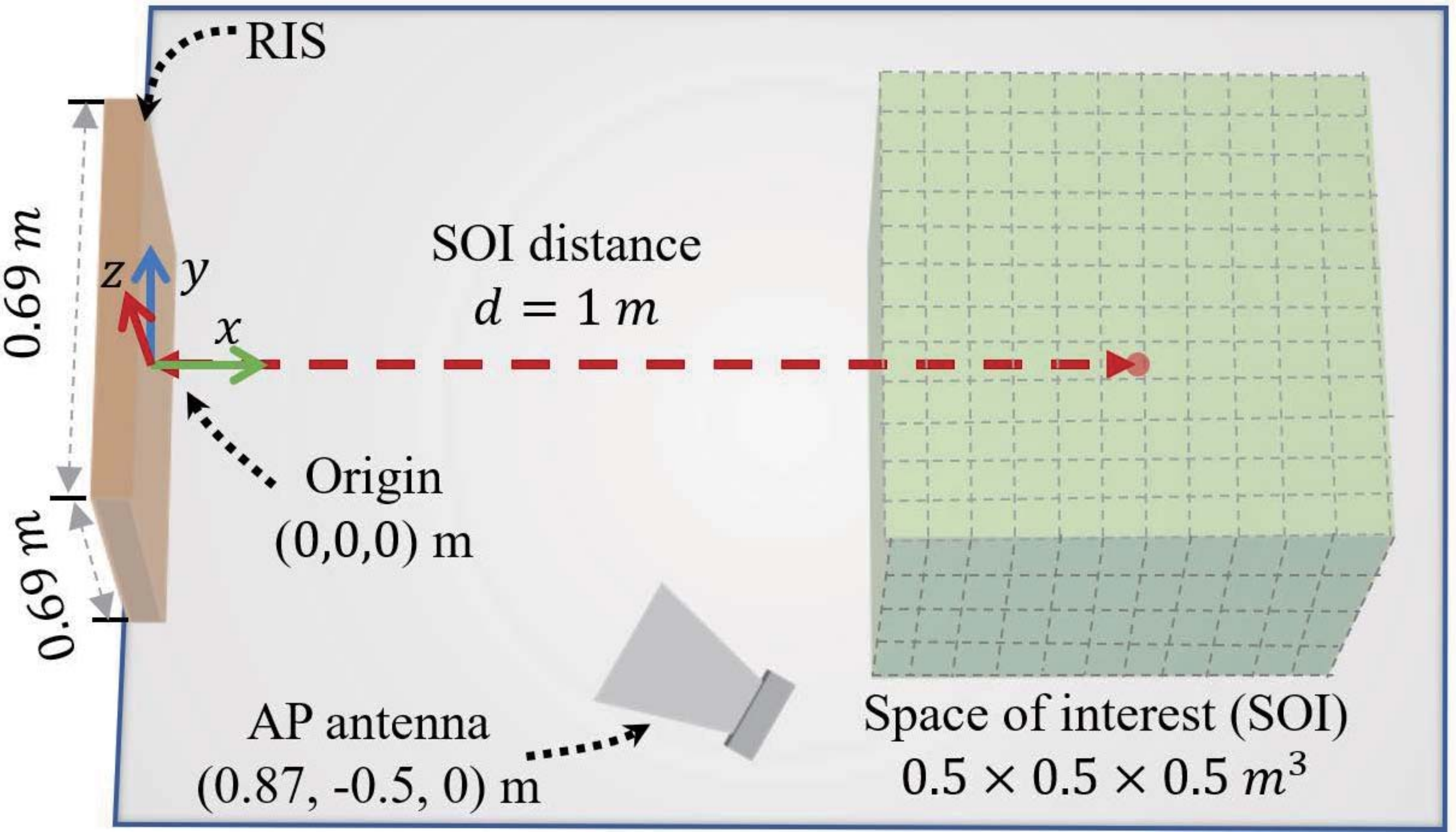}
	\caption{Experimental layout for the MetaLocalization system.}
	\label{f_layout}
\end{figure}

\subsubsection{RIS-aided Localization}
For localization applications, we show the MetaLocalization system as an example. We perform the experiments in a classroom with the size of 25 m$^2$, and the walls of the classroom are made of bricks and concrete. As illustrated in Fig.~\ref{f_layout}, the SOI is a cuboid region with the size of $0.5 \times 0.5 \times 0.5$m$^3$. The center of the SOI is 1 m away from the center of the RIS. When building the radio map for each SOI, we discretize the SOI into blocks and record the signals for each block. No objects exist between the users and the RIS. Please refer to \cite{HJHBKZL-2021} for more experimental details.

\begin{figure*}[!t]
	\centering
	\subfloat[]{
		%\label{a1} %% label for first subfigure
		\includegraphics[width=2.35in]{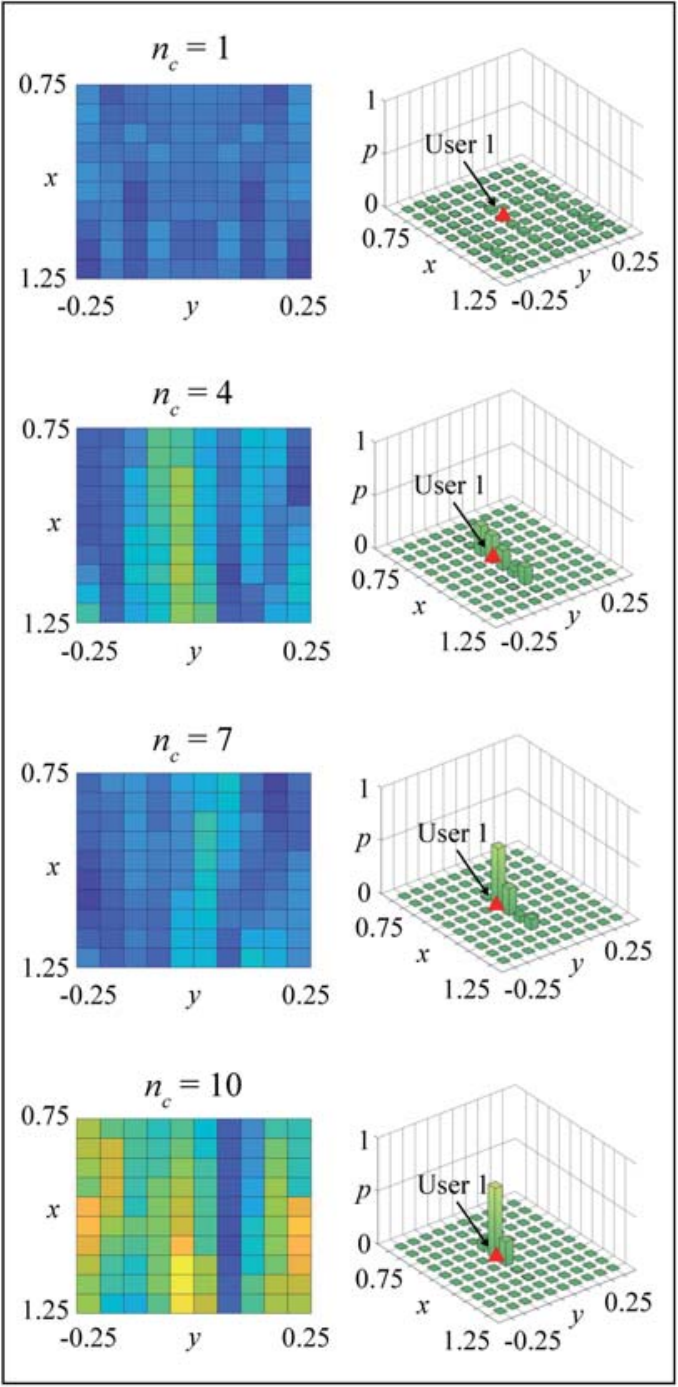}}
	%\hspace{0.005in}
	\subfloat[]{
		%\label{a1} %% label for first subfigure
		\includegraphics[width=2.35in]{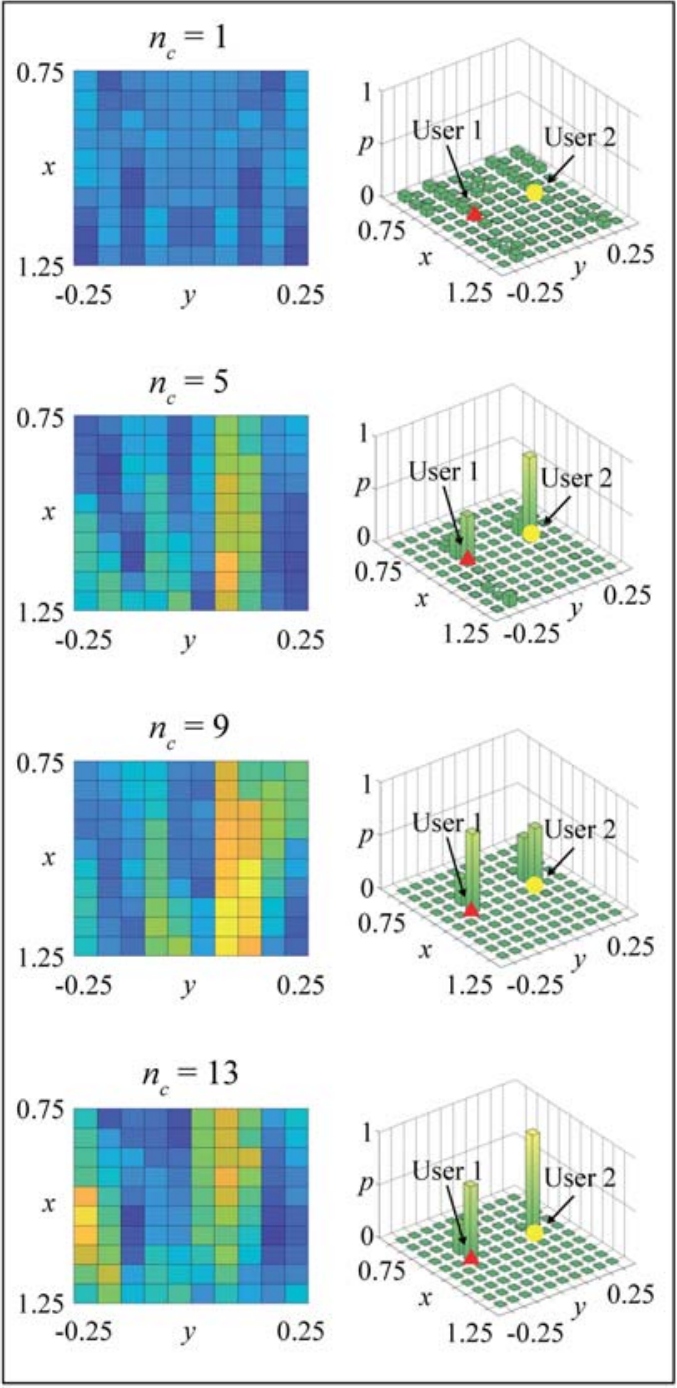}}
	%\hspace{0.005in}
	\subfloat[]{
		%\label{a1} %% label for first subfigure
		\includegraphics[width=2.35in]{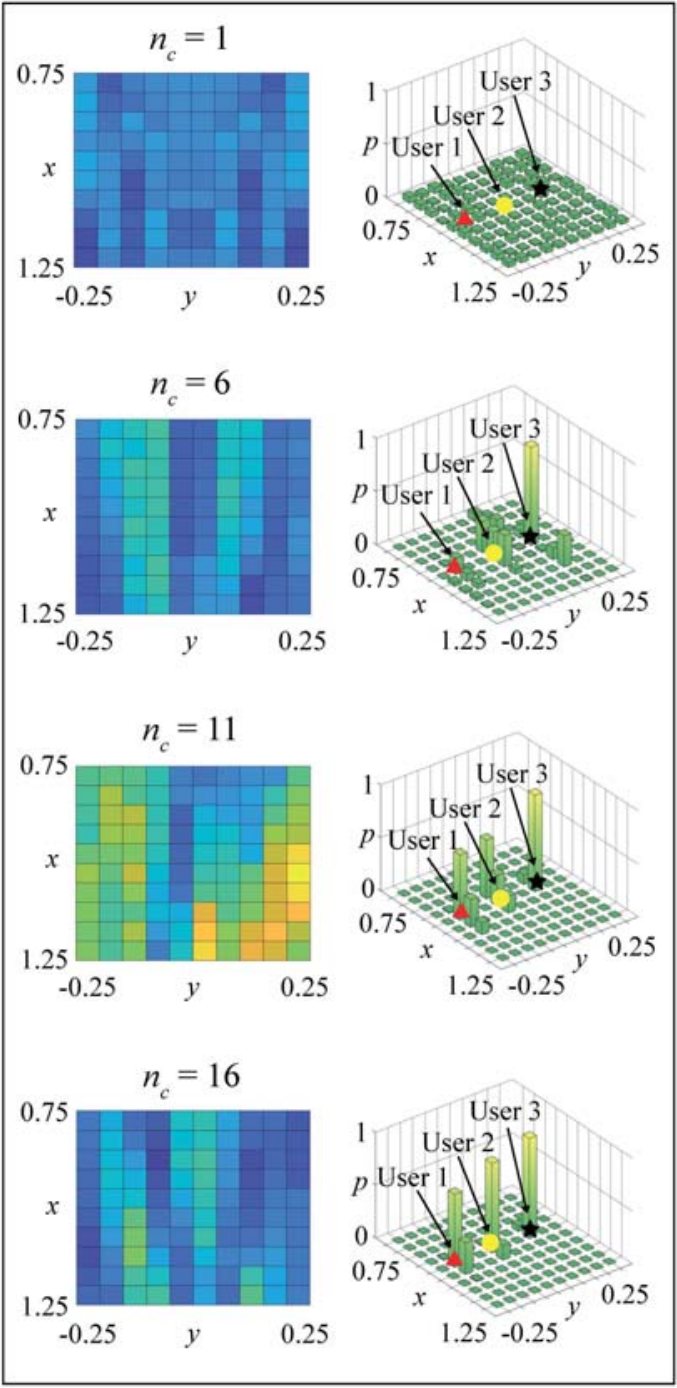}}
	
	\caption{Illustrations of the Metalocalization scheme without obstruction for three users: (a) One user; (b) Two users; (c) Three users. The first column in each subfigure shows the radio maps after different cycles, and the second column shows the corresponding probability distribution for different locations. Here, the probability is the sum of the probabilities of all the users. The ground truth of three users' locations are denoted by the red triangle, yellow circle, and the black star, respectively.}
	\label{f_example}
\end{figure*}

Fig.~\ref{f_example} illustrates the process of multi-user localization without obstruction. For display simplicity, in this figure, we choose a planar SOI which is on the plane $z = 0$ with size $0.5 \times 0.5$m$^2$ and the distance to the RIS is $1$m. The ground truth of users' locations are labeled in the figures by the red triangle (the first user), yellow circle (the second user), and the black star (the third user), respectively. We can observe the RSS varies for different iterations. The probabilities are approximately uniformly distributed in each location in the first cycle (first row of subfigures), while after several cycles, the probabilities of locations near the ground truth are obviously higher than those in the other locations, which implies the effectiveness of the MetaLocalization scheme. We can also observe that the locations in the $x$ direction near the ground truth have higher probabilities than the locations in the $y$ direction, indicating that it is more likely to misjudge the $x$ coordinate than the $y$ coordinate of the user's location. The reason for this observation is that the correlation of signals in the $x$ axis is higher than those in the $y$ and $z$ directions as the $x$ axis is perpendicular to the RIS.

\section{Future Directions}
\label{Future}
In previous sections, we have presented some case studies to illustrate how to integrate the RIS to wireless networks to achieve ubiquitous sensing and localization, and have introduced how to build the prototypes. In the following, we will further discuss some other relevant topics that are worthy of further investigation.

\subsection{Mobility}
In the previous case studies, we focus on the static or slow moving targets where the targets/users stay in the same block within a cycle. However, in practical scenarios, the targets/users might move with a relatively high speed, e.g., vehicles, which implies that they will be in different blocks within a cycle. Therefore, it is also necessary to develop new techniques for such use cases by exploiting the relations among these blocks.

Different from the traditional methods for moving targets/users \cite{CZZXYJ-2015}, where the motion is purely captured by the signal processing techniques at the Rx, the RIS-aided localization and sensing systems can leverage the capability to configurate the RIS based on the motion of the targets to achieve higher sensing or localization accuracy. However, if we purely configurate the RIS based on the predicted locations, the estimation errors in previous slots might accumulate, leading to a low received signal strength from the objects. As a result, \emph{how to develop an effective RIS configuration scheme according to the movement of objects still remains an open problem.} On the other hand, the signal processing techniques at the Rx will also be different. The RIS configuration also influences the received signal strength and should be jointly designed with the estimation function. Therefore, \emph{how to estimate a series of locations of a moving targets is a challenge.}

\subsection{Millimeter-wave and Terahertz Bands}
The higher frequency bands, including millimeter-wave (mmWave) and terahertz (THz) bands, could be promising candidates for sensing and localization applications as they can provide fine resolution in range and angle. However, the higher frequency bands suffer high propagation losses and power limitations results in a short sensing and localization range \cite{HNTM-2020}. Benefited from the capability to customize the propagation environment, the RIS has shown its potential to improve the range, but the rendezvous of RISs and mmWave and THz bands also brings some unique challenges for RF sensing and localization use cases.
\begin{itemize}
	\item \textbf{Hardware Complexity:} The RIS is typically designed to operate over a predefined frequency band, and the implementation complexity is positively proportional to the working frequency as the size of the RIS element should be on the order of the wavelength. Moreover, it also requires faster response time for changing phase shifts.
	
	\item \textbf{Signal Processing:} Higher frequency bands have some unique characteristics, which require to develop some signal processing techniques specific to these bands. For example, the path loss of a THz signal in the presence of water vapor is dominated by spikes that represent molecular absorption losses due to excited molecule vibrations \cite{HMT-2021}. As a result, the spectrum is divided into small subbands and these subbands are distance-dependent. Therefore, it is important to consider these physical features when designing the decision function for sensing and localization purposes.
\end{itemize}

\subsection{Security and Privacy}
The fast development of wireless and localization technologies have led to the flourish of location-based services (LBSs), where delivered information is customized according to users' physical locations. Although LBSs can provide enhanced functionalities, they also open up new vulnerabilities that can be used to cause security and privacy issues \cite{MLJHSMZ-2022}. For example, location information privacy is becoming critical in various applications, such as health monitoring, social media, and surveillance systems, as we can infer some sensitive information from users' locations. Therefore, it is important to preserve users' privacy and security.

In the RIS-aided localization systems, the adoption of the RIS also brings some unique challenges and research directions for users' privacy and security preserving as listed below.
\begin{itemize}
\item \textbf{Privacy Preserving Mechanisms:} In traditional location-based systems, there have existed some privacy preserving mechanisms, such as mix-zones, dummies-based mechanisms, and perturbation-based mechanisms \cite{VASL-2019}. However, these mechanisms cannot be applied in RIS-aided localization systems since we might also infer users' locations from the configurations of the RIS. Therefore, it is necessary to consider the trade-off between the accuracy and privacy when optimizing the configuration of the RIS.

\item \textbf{Secure Localization Methods:} Although the RIS provides the opportunity to customize the propagation environment, it will also become an attack target leading to new security issues. For example, if the RIS is attacked and its configuration is not consistent with that computed at the BS, the estimated location will be quite different from the ground-truth one, which might cause severe consequences in many critical applications. Therefore, it is also important for the RIS-assisted localization systems to detect the misbehavior of the RIS.

\end{itemize}

\subsection{Integrated Sensing and Communication Design}
In general, sensing and communication systems are deployed separately and use different frequency bands. However, with the increasing number of connected devices and services in wireless communication industry, the frequency is becoming congested. This motivates us to consider the integration of sensing and communication to further improve the spectrum efficiency \cite{FCAPHL-2020}. As a result, the integrated sensing and communication (ISAC) system needs to be well designed to achieve the performance trade-off between sensing and communication. 

In the RIS-aided sensing systems, it is not easy to integrate both sensing and communication functions. Some possible challenges and research directions are listed as follows.
\begin{itemize}
	\item \textbf{Channel Modeling:} The algorithm for sensing functions highly depends on the location of targets and surrounding environment. This cannot be captured by stochastic channel models that is widely used for wireless communications. To this end, ray tracing could be a strong candidate for channel modeling methodology \cite{DJYAYPW-2021}. However, the existence of the RIS makes the ray tracing more complicated as the RIS might introduce the multi-hop scattering. Therefore, it is essential to develop a new channel modeling method to accommodate the RIS assisted ISAC system.
	
	\item\textbf{Waveform Design:} In the ISAC system, a single RF signal should convey both communication and sensing data. Therefore, the waveform design is important but challenging due to the contradicting metrics for the communication and sensing. In particular, the main target for communication systems is to maximize the spectral efficiency \cite{XBHCCZ-2021}, while the optimal waveform for the sensing function is designed for a higher sensing resolution and accuracy. Therefore, lot of effort needs to be made for the waveform design to strike a balance between the communication and sensing performance.
	
	\item\textbf{RIS Configuration:} The deployment of the RIS in an ISAC system can shape the radio environment by adjusting the phase responses of each RIS element in order to improve both sensing and communication performance. This requires an appropriate design for the configuration of the RIS. However, the optimization of the RIS configuration is not trivial. First, in practical systems, the number of possible phase shifts applied to each RIS element is finite \cite{HBLZ-2020}. As a result, the feasibility set for the optimization problem is discrete, leading to an NP-hard integer program. Second, the RIS configuration is, in general, coupled with the waveform design, which makes it more complicated. Machine learning methods might be a powerful tool to address this issue. Moreover, how to place the RIS to obtain a better performance is another interesting topic \cite{SHBZL-2021}.  
\end{itemize}

\section{Conclusions}
\label{Conclusion}
In this paper, we have provided a comprehensive tutorial on the application of the innovative RIS technique to wireless sensing and localization use cases. Benefited from its capability to customize the wireless propagation environment, the RIS has shown the potential to enhance the difference of the received signals from neighboring targets/locations, and thus improving the accuracy of sensing and localization. This paper introduces the preliminaries, state-of-the-art results on main challenges for wireless sensing and localization applications, and future research directions. We hope that this paper can be a useful resource for future research on RIS-aided wireless sensing and localization, unlocking its full potential in future cellular systems.

\end{document}